\documentclass[english]{svjour3}
\usepackage[latin9]{inputenc}
\usepackage{color}
\usepackage{url}
\usepackage{amsmath}
\usepackage{graphicx}
\usepackage{wasysym}
\PassOptionsToPackage{normalem}{ulem}
\usepackage{ulem}

\makeatletter

\providecommand{\tabularnewline}{\\}

\usepackage[T1]{fontenc}

\usepackage{amsmath}
\usepackage{amssymb}
\usepackage{wasysym}

\usepackage[sort&compress,numbers]{natbib}
\newcommand{\E}{\mathrm{e}}

\usepackage[unicode=true,
 bookmarks=false,
 breaklinks=false,
 colorlinks=true,
 allcolors=blue]
 {hyperref}

\usepackage{microtype}

\makeatother

\begin{document}
\title{Beyond just ``flattening the curve'': Optimal control of epidemics
with purely non-pharmaceutical interventions}
\titlerunning{Beyond just ``flattening the curve'': Optimal control of epidemics
...}
\author{Markus Kantner and Thomas Koprucki}
\authorrunning{M. Kantner and T. Koprucki}
\institute{M. Kantner~$\cdot$~T. Koprucki\\
 Weierstrass Institute for Applied Analysis and Stochastics (WIAS)\\
 Mohrenstr. 39, 10117 Berlin, Germany\\
 \email{kantner@wias-berlin.de, koprucki@wias-berlin.de}}
\date{}
\date{\today \qquad(First version: April 17, 2020)}
\maketitle
\begin{abstract}
When effective medical treatment and vaccination are not available,
non-pharmaceutical interventions such as social distancing, home quarantine
and far-reaching shutdown of public life are the only available strategies
to prevent the spread of epidemics. Based on an extended SEIR (susceptible-exposed-infectious-recovered)
model and continuous-time optimal control theory, we compute the optimal
non-pharmaceutical intervention strategy for the case that a vaccine
is never found and complete containment (eradication of the epidemic)
is impossible. In this case, the optimal control must meet competing
requirements: First, the minimization of disease-related deaths, and,
second, the establishment of a sufficient degree of natural immunity
at the end of the measures, in order to exclude a second wave. Moreover,
the socio-economic costs of the intervention shall be kept at a minimum.
The numerically computed optimal control strategy is a single-intervention
scenario that goes beyond heuristically motivated interventions and
simple ``flattening of the curve''. Careful analysis of the computed
control strategy reveals, however, that the obtained solution is in
fact a tightrope walk close to the stability boundary of the system,
where socio-economic costs and the risk of a new outbreak must be
constantly balanced against one another. The model system is calibrated
to reproduce the initial exponential growth phase of the COVID-19
pandemic in Germany.


\renewcommand{\subclassname}{\textbf{MSC (2010)}\;\;}
\end{abstract}

\vspace{-2.5ex}

\section{Introduction \label{sec:Introduction}}

\noindent Preventing the spread of new diseases, to which there is
no immunity in the population, is a huge problem, since there are
often neither vaccines nor other effective medical treatments available
in the early stages. In this case, non-pharmaceutical interventions
(NPIs) such as intensive hand hygiene, home quarantine and measures
of social distancing, e.g. closure of schools, universities and shops,
prohibition of mass events up to curfew and shutdown of entire territories,
are the only available measures. The NPIs are aimed at ``flattening
the curve'', i.e., a reduction of the transmission rate in order
to break the exponential growth of the epidemic.

In the case of the currently spreading COVID-19 pandemic caused by
the new SARS-CoV2 coronavirus \cite{Zhu2020,Wu2020}, the fundamental
concern of the mitigation measures is not to exceed the available
number of intensive care unit (ICU) beds, in particular for respiratory
support or extracorporeal membrane oxygenation, in order to prevent
actually avoidable deaths \cite{Phua2020}. Since the outbreak of
the epidemic, a large number of simulation studies have been conducted
using mathematical models to assess the efficacy of different NPIs
and to estimate the corresponding demands on the health care system
\cite{Ferguson2020,Flaxman2020,Chang2020,Ng2020,Bouchnita2020,Jia2020,Kucharski2020,Barbarossa2020,Kissler2020}.
Moreover, mathematical models are employed to deduce important epidemiological
parameters \cite{Sesterhenn2020,Khailaie2020,Engbert2020} and to
evaluate the effect of particular measures from empirical data \cite{Dehning2020,Brauner2020}.

The vast majority of research papers on the control of COVID-19 examines
the impact of rather simple intervention schemes such as bang-bang
control or cascaded on-off (i.e., repeated lockdown and release) strategies
\cite{Kissler2020,Tsay2020,Tarrataca2020,Bin2020}. Instead, however,
intervention strategies derived from continuous-time optimal control
theory \cite{Lewis2012} following a variational principle are actually
preferable. There is a large number of studies on the application
of optimal control theory following Pontryagin's maximum principle
\cite{Pontryagin1962} in mathematical epidemiology, see Refs.~\cite{Wickwire1977,Sharomi2015,Behncke2000,Nowzari2016,Lenhart2007}
and references therein. The by far largest part of these works deals
with optimal control of epidemics through vaccination and immunization
\cite{Morton1974,Abakuks1974,Zaman2008,Kar2011}, medical treatment
\cite{Zaman2009,Liddo2016} and combinations thereof \cite{Gaff2009,Hansen2011,Iacoviello2013,Bolzoni2017,Barro2018,Bolzoni2019}.
Significantly fewer papers are concerned with the optimal control
of transmission dynamics and the mitigation of epidemics through social
distancing measures. The paper by Behncke \cite{Behncke2000} studies
the optimal control of transmission dynamics via optimally steered
health-promotion campaigns and seems to be one of the first works
devoted to this problem. During the current COVID-19 pandemic, the
control of the disease by NPIs has moved into the focus of attention
and a number of recent papers are devoted to this problem. Djidjou-Demasse
et al. \cite{Djidjou-Demasse2020} investigated the optimal control
of the epidemic via social distancing and lockdown measures until
a vaccine becomes available. They propose to delay the peak of the
epidemic by increasingly strict interventions and finally to relax
the measures in such a way that a significant burden on the health
care system only occurs when the availability of a vaccine is already
expected. A similar problem has been considered by Perkins and Espa\~{n}a
\cite{Perkins2020}, who studied the optimal implementation of NPIs
under the assumption that an effective vaccine would become available
in about one year after the outbreak of the epidemic. The paper by
Kruse and Strack \cite{Kruse2020} is devoted to the analysis of the
optimal timing of social distancing measures under the constraint
that the overall (temporal) budget for NPIs is limited. Ketcheson
\cite{Ketcheson2020} presented a detailed analysis for optimal transmission
control in a SIR (susceptible-infected-recovered) epidemic model with
the aim of achieving a stable equilibrium (``herd immunity'') within
a fixed finite time interval while simultaneously avoiding hospital
overflow. A similar problem (including a simple state-dependent mortality
rate) was studied by Alvarez et al. \cite{Alvarez2020}, who focussed
on minimizing the lockdown costs and included further economic aspects
such as the assumed value of statistical life. An extension of the
optimal transmission control problem to an age-structured model has
been presented by Bonnans and Gianatti \cite{Bonnans2020}, who proposed
a different temporal course of the contact reduction for the high
and low risk sub-populations. Köhler et al. \cite{Koehler2020} have
applied model predictive control to social distancing measures with
the objective of minimizing the fatalities over a fixed period of
time of two years. Next to adaptive feedback strategies for iterative
loosening of the social distancing policies after an initial lockdown,
the authors also examined the possibility of eradicating the virus.
All of these papers on optimal control deal with deterministic epidemiological
models, in particular the basic SIR model \cite{Kruse2020,Behncke2000,Miclo2020,Alvarez2020,Ketcheson2020}
or various extended SEIR-type models \cite{Djidjou-Demasse2020,Perkins2020,Koehler2020}.
We remark that this survey on optimal control of COVID-19 is not exhaustive.

The objective of this paper is the investigation of the optimal control
of epidemics in the (hopefully unlikely) case in which an effective
vaccine is impossible or never found and the epidemic must be controlled
with purely non-pharmaceutical measures. Furthermore, we exclude the
possibility of complete containment (``eradication of the virus'').
Then, optimal control must pursue competing objectives: On the one
hand, the number of disease-related deaths shall be minimized by strictly
avoiding an overload of the intensive care treatment capacities. On
the other hand, however, sufficient natural immunity must be established
in the population in the long run to prevent a second outbreak of
the epidemic (``herd immunity''). Moreover, the socio-economic costs
of the intervention shall be kept at a minimum. We compute the optimal
solution to this problem by applying Pontryagin's Maximum Principle
to an extended SEIR-type model tailored to specific aspects of COVID-19.
Our main result is the optimal time course of the mean contact reduction
(and the corresponding time-dependent effective reproduction number)
that serves as a guideline on how to optimally enter and finally exit
the lockdown. The corresponding NPI policy is a single-intervention
scenario that can be divided into three distinct phases: (1)~a strict
initial lockdown, (2)~a long lasting period (``critical period'')
during which the number of active cases is kept approximately constant
and (3)~a moderate tightening of the measures towards the end of
the intervention. We present a detailed analysis of the numerically
computed result and develop an analytical understanding of its distinct
features. Moreover, we show that our numerically computed optimal
control obeys two fundamental stability criteria, which impose an
upper limit on the transmission rate and its rate of change on the
way out of the initial lockdown. The precise structure of the optimal
control (i.e., three phases of the intervention) obtained in this
paper differs from the results described in similar works \cite{Alvarez2020,Kruse2020,Ketcheson2020}.
After the initial submission of this paper, the preprint by Charpentier
et al. \cite{Charpentier2020} appeared, who studied a similar optimization
problem on the basis of an extended SIR-type model with parameters
adjusted to the COVID-19 pandemic in France. Their independently obtained
results are comparable to those presented in this paper, which demonstrates
the robustness of the obtained optimal intervention strategy with
respect to model and parameter variations.

The mathematical model for the progression of the epidemic and the
estimation of the demand for intensive care resources is described
in Sec.~\ref{sec:Mathematical-modeling}. The optimal control problem
is derived in Sec.~\ref{sec:Optimal-control} and the results are
described in Sec.~\ref{sec:Results}. We close with a critical discussion
of our findings in Sec.~\ref{sec: Summary-and-Conclusions}. The
model has been calibrated to reproduce the exponential growth phase
of the COVID-19 pandemic in Germany. Details on the parameter adjustment
are described in the Appendix.

\section{Modeling of Disease Spreading and Demand for Intensive Care Units
\label{sec:Mathematical-modeling}}

Mathematical modeling of the spread of epidemics is an indispensable
tool to project the outcome of an epidemic, estimate important epidemiological
parameters and to make predictions for different intervention scenarios.
Compartment models \cite{Kermack1927,Hethcote2000,Brauer2008}, where
the population is divided into different macroscopic sub-populations,
such as \emph{susceptible}, \emph{infectious}, \emph{recovered} etc.,
are a simple but effective tool to model the progression of epidemics.
In contrast to complex (but more realistic) stochastic agent-based
models \cite{Epstein2009,Rahmandad2008}, deterministic mean-field
models are limited to the description of the average infection dynamics
in macroscopic (sub-)populations, but allow for fast parameter scans
and a straightforward application of continuous-time optimal control
theory \cite{Lewis2012}.

\begin{figure}[t]
\includegraphics[width=1\textwidth]{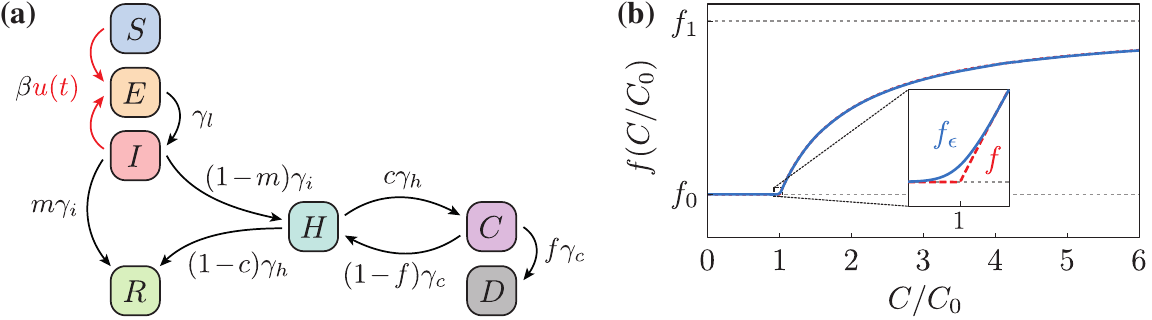}\caption{\textbf{(a)}~Schematic illustration of the compartmental epidemic
model (\ref{eq: model equations}). The function $u\left(t\right)$
describes a modification of the transmission dynamics due to NPIs.
\textbf{(b)}~State-dependent mortality rate $f$ as a function of
the number of patients in a critical state requiring intensive care.
The mortality rate grows rapidly if the number of critical patients
exceeds the number of available ICUs $C_{0}$. Inset: The solid line
is the regularized mortality rate (\ref{eq: fatality (smooth)}) that
is used in the computations throughout the paper.}

\label{fig: schematic and f}
\end{figure}

\subsection{Model equations}

In this paper, an extended SEIR model, similar to that proposed by
Neher et al. \cite{Neher2020,Noll2020}, is used to model the spread
of an epidemic and to estimate the number of patients in a critical
state that require intensive care. Similar models are described in
Refs.~\cite{Khailaie2020,Wilson2020,Koehler2020}. For the sake of
simplicity, vital dynamics (except for disease-related deaths), seasonality
effects \cite{Neher2020a}, dispersion of transmission \cite{LloydSmith2005}
and any effects caused by population heterogeneity (different age
and risk groups) are neglected. The total population is divided into
distinct compartments: susceptible $S$, exposed $E$, infectious
$I$, hospitalized $H$ (severely ill), critical $C$, recovered $R$
(i.e., immune) and deceased $D$. The model equations read\begin{subequations}\label{eq: model equations}
\begin{align}
\dot{S} & =-\beta u\left(t\right)\frac{IS}{N},\label{eq: model equations - S}\\
\dot{E} & =\beta u\left(t\right)\frac{IS}{N}-\gamma_{l}E,\label{eq: model equations - E}\\
\dot{I} & =\gamma_{l}E-\gamma_{i}I,\label{eq: model equations - I}\\
\dot{H} & =\left(1-m\right)\gamma_{i}I+\left(1-f\left(C/C_{0}\right)\right)\gamma_{c}C-\gamma_{h}H,\label{eq: model equations - H}\\
\dot{C} & =c\gamma_{h}H-\gamma_{c}C,\label{eq: model equations - C}\\
\dot{R} & =m\gamma_{i}I+\left(1-c\right)\gamma_{h}H,\label{eq: model equations - R}\\
\dot{D} & =f\left(C/C_{0}\right)\gamma_{c}C.\label{eq: model equations - D}
\end{align}
\end{subequations}The group of initially healthy and not yet infected
(susceptible, $S$) is vulnerable to infection through contact with
infectious ($I$), who may transmit the disease to the susceptible
population. The infection probability is determined by the transmission
rate $\beta$, and the share of the susceptible and infectious population
on the total (living) population $N=N(t)$, which is given as
\begin{equation}
N=S+E+I+H+C+R.\label{eq: total population}
\end{equation}
The newly infected (exposed, $E$) become infectious themselves only
after a latency period $\gamma_{l}^{-1}$ (which must not be confused
with the incubation time). The infectious either recover or turn severely
ill after an average period $\gamma_{i}^{-1}$. Severely ill ($H$)
can either deteriorate into a critical state ($C$) or recover after
a period $\gamma_{h}^{-1}$. The recovered population ($R$) is assumed
to be immune against new infections. Patients in a critical state
either stabilize to the severely ill state or die from the disease
on a time scale $\gamma_{c}^{-1}$. The disease-related deaths reduce
the size of the population
\begin{equation}
\dot{N}=-\dot{D},\label{eq: population dynamics}
\end{equation}
such that, assuming initially $D\left(0\right)=0$, it holds $N\left(t\right)=N\left(0\right)-D\left(t\right)$.
Moreover, $m$ is the share of infectious that are asymptomatic or
have at most mild symptoms, $c$ is the fraction of severely ill that
become critical and $f$ is the fraction of critically ill that are
going to die from the disease. Finally, the time-dependent function
$u\left(t\right)$ describes a modification of the transmission rate
(mean contact reduction) due to NPIs. Here, $u=1$ means no intervention,
and $u=0$ corresponds to the extreme case of total isolation of the
whole population. The model system is illustrated in Fig.~\ref{fig: schematic and f}\,(a).
A rescaled version of the dynamical system \eqref{eq: model equations},
where the sub-populations are considered in terms of their relative
share of the initial population $N\left(0\right)$, is given in Eq.~\eqref{eq: model equations - rescaled}
in the Appendix~\ref{sec: Two-point BVP and initial values}.

\begin{figure}
\includegraphics[width=1\textwidth]{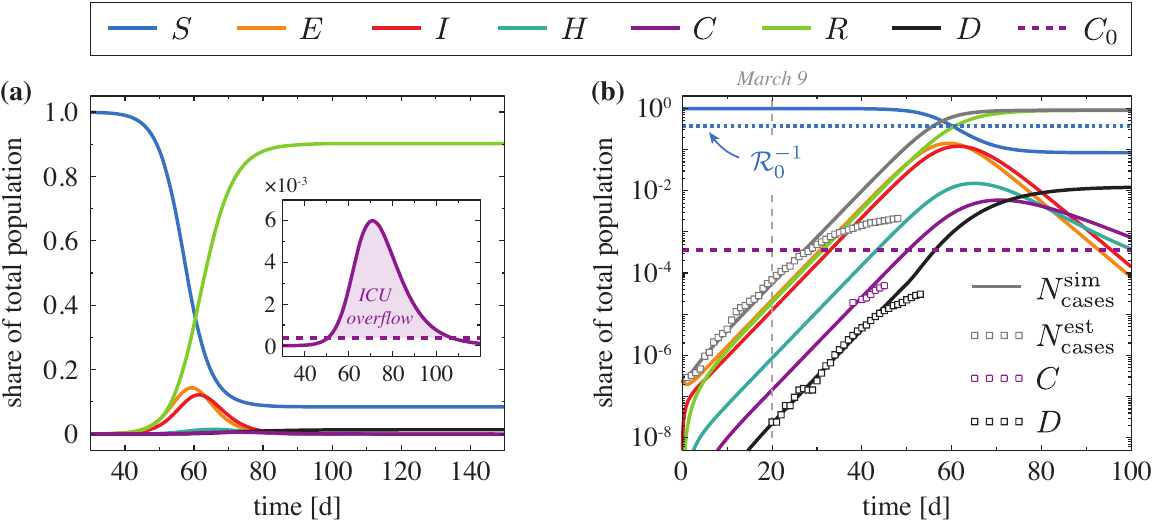}

\caption{\textbf{(a)~}Evolution of the epidemic without interventions ($u=1$).
The number of available ICUs was set to $C_{0}=30\,000$. The inset
shows the overflow in ICU demand, which leads during a period of about
57 days to an increased mortality rate according to Eq.~(\ref{eq: fatality}).
\textbf{(b)}~Same as in (a) but on a logarithmic scale. The markers
indicate the estimated number of cumulative cases (see Appendix \ref{sec:Parameter-adjustment})
and the reported numbers for ICU demand and deaths during the early
phase of the COVID-19 pandemic in Germany. The first disease-related
fatalities were reported on March 9, 2020 (day number 20 in the simulation).
Social distancing measures, which came into force nationwide in mid-March
\cite{Dehning2020}, have flattened the initial exponential growth.}

\label{fig: uncontrolled}
\end{figure}

\subsection{State-dependent fatality rate}

The disease-related mortality grows tremendously as soon as the number
of critically ill exceeds the capacity limit $C_{0}$ of the health
care system (number of available ICUs). This is modeled by a state-dependent
average fatality rate\begin{subequations}\label{eq: fatality} 
\begin{equation}
f=f\left(\frac{C}{C_{0}}\right)=\begin{cases}
f_{0} & \text{for }C\leq C_{0},\\
f_{1}-\frac{C_{0}}{C}\left(f_{1}-f_{0}\right) & \text{for }C>C_{0}.
\end{cases}\label{eq: fatality (non-differentiable)}
\end{equation}
As long as every critical patient can be served with an ICU ($C\leq C_{0}$),
the fatality rate is a constant $f=f_{0}$. As soon as the ICU resources
are exceeded, an increasing fraction of the critical patients dies
with a higher rate $f_{1}>f_{0}$, which on average results in the
state-dependent fatality rate~(\ref{eq: fatality (non-differentiable)}).
Here, $f_{1}=2f_{0}$ is assumed. In the following, the regularization
\begin{equation}
f\left(x\right)\to f_{\epsilon}\left(x\right)=f_{0}+\frac{\epsilon}{x+\ensuremath{1.1}\epsilon}\log{\left(1+\exp{\left(\frac{x-1}{\epsilon}\right)}\right)}\left(f_{1}-f_{0}\right)\label{eq: fatality (smooth)}
\end{equation}
\end{subequations}with $0<\epsilon\ll1$, of Eq.~(\ref{eq: fatality (non-differentiable)})
is used, in order to avoid problems due to the non-differentiability
at $C=C_{0}$. The function $f\left(C/C_{0}\right)$ is plotted in
Fig.~\ref{fig: schematic and f}\,(b).\sout{}

\begin{table}
\centering

\begin{tabular}{ccl}
\hline 
symbol & value & description\tabularnewline
\hline 
$\mathcal{R}_{0}$ & $2.7$ & basic reproduction number\tabularnewline
$N(0)$ & $83\times10^{6}$ & initial population size\tabularnewline
$\gamma_{l}^{-1}$ & $2.6\,\text{d}$ & average latency time between exposure and infectious period\tabularnewline
$\gamma_{i}^{-1}$ & $2.35\,\text{d}$ & average infectious period before recovery or hospitalization\tabularnewline
$\gamma_{h}^{-1}$ & $4.0\,\text{d}$ & average period before severely ill patients turn critical or recover\tabularnewline
$\gamma_{c}^{-1}$ & $7.5\,\text{d}$ & average period before critical patients recover or die\tabularnewline
$\beta$ & $\left(1.15\,\text{d}\right)^{-1}$ & transmission rate\tabularnewline
$m$ & $0.92$ & fraction of infected with at most mild symptoms\tabularnewline
$c$ & $0.27$ & fraction of hospitalized patients that turn critical\tabularnewline
$f$ & see Eq.~(\ref{eq: fatality}) & fraction of critical patients that turn fatal\tabularnewline
$f_{0}$ & $0.31$ & mortality of a critical patient with ICU\tabularnewline
$f_{1}$ & $2\thinspace f_{0}=0.62$ & mortality of a critical patient without ICU\tabularnewline
$C_{0}$ & variable & number of ICUs/ max. number of simultaneously critical cases\tabularnewline
$T$ & $10\times T_{\text{crit}}$ & final time of the simulation, for $T_{\text{crit}}$ see Eq.~(\ref{eq: Tcrit})\tabularnewline
\hline 
\end{tabular}

\caption{List of parameters used in the simulations. See Appendix \ref{sec:Parameter-adjustment}
for details.}

\label{tab: parameters}
\end{table}

\subsection{Basic and effective reproduction number}

The basic reproduction number \cite{Diekmann1990}
\begin{equation}
\mathcal{R}_{0}=\beta/\gamma_{i}\label{eq: basic reproduction number}
\end{equation}
can be thought of as the expected number of cases (without intervention,
$u=1$) that is directly generated by one case in a population where
all individuals are susceptible to infection. The effective reproduction
number 
\begin{equation}
\mathcal{R}_{\text{eff}}\left(t\right)=\mathcal{R}_{0}u\left(t\right)S\left(t\right)/N\left(t\right)\label{eq: effective reproduction number}
\end{equation}
depends on time and includes the impact of intervention measures.

\subsection{Numerical results for the uncontrolled epidemic (COVID-19 in Germany)}

Figure~\ref{fig: uncontrolled} shows the progression of an uncontrolled
epidemic starting from an initially small fraction of exposed population.
The initial conditions are listed in Appendix~\ref{sec: Two-point BVP and initial values}.
The parameters are adjusted (see Appendix~\ref{sec:Parameter-adjustment})
to reproduce the initial exponential growth phase of the COVID-19
disease in Germany (late February -- mid March 2020) and are summarized
in Tab.~\ref{tab: parameters}. The numerical solution was obtained
by a 4th order Runge--Kutta method. Without intervention, the peak
number of simultaneously active cases is about $23$ million and the
peak number of patients in a critical state exceeds the number of
ICUs by a factor of about $C_{\text{max}}/C_{0}\approx16.7$, see
inset of Fig.~\ref{fig: uncontrolled}\,(a). The simulated value
$C_{\text{max}}\approx5.0\times10^{5}$ is in very good agreement
with the projection by Khailaie et al. \cite{Khailaie2020}. Due to
the increased fatality in the period with ICU overflow, see Eq.~(\ref{eq: fatality}),
the epidemic terminates with a very high number of deaths $D\left(T\right)\approx1.0\times10^{6}$,
which is in line with previous studies \cite{Barbarossa2020}.

\section{Optimal Control \label{sec:Optimal-control}}

\begin{figure}
\sidecaption\includegraphics{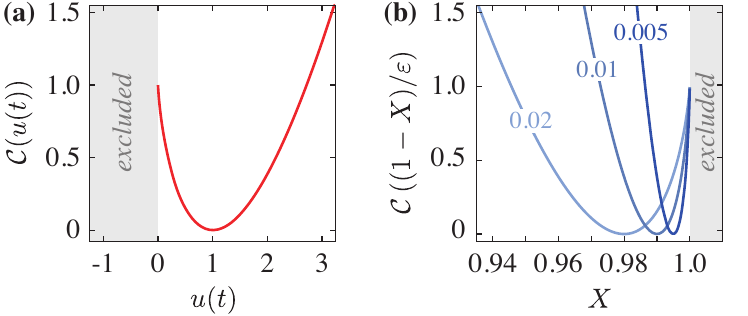}\hfill\caption{Plot of the cost functions for \textbf{(a)~}minimal intermediate
costs and \textbf{(b)}~the enforcement of herd immunity at the end
of the intervention for different values of $\varepsilon$. We use
the short notation $X=\mathcal{R}_{0}S\left(T\right)/N\left(T\right)$.
The shaded region corresponds to unstable terminal states.\vspace{2.95ex}}
\label{fig: cost functions}
\end{figure}

In the scenario outlined in Sec.~\ref{sec:Introduction}, where an
effective vaccine is never found, the optimal transmission control
due to NPIs is required (i)~to avoid ICU overflow (more patients
in a critical state than available ICUs) but at the same time (ii)~exclude
a second wave of the epidemic after the end of the measures.
The optimal solution is computed by minimizing the index functional\begin{subequations}\label{eq: index functional - total}
\begin{equation}
\mathcal{J}\left[u\right]=\varphi\left(\mathbf{x}\left(T\right)\right)+\int_{0}^{T}\mathrm{d}t\,\mathcal{C}\left(u\left(t\right)\right)\label{eq: index functional}
\end{equation}
where
\begin{equation}
\varphi\left(\mathbf{x}\left(T\right)\right)=\mathcal{P}\,D\left(T\right)+\mathcal{C}\left(\frac{1}{\varepsilon}\left(1-\mathcal{R}_{0}\frac{S\left(T\right)}{N\left(T\right)}\right)\right)\label{eq: terminal cost function}
\end{equation}
\end{subequations}is the terminal cost function. The first term in
Eq.~(\ref{eq: terminal cost function}) describes the number of disease-related
deaths $D\left(T\right)$ at the end of the epidemic, which should
be minimized. As the increment of the disease-related deaths depends
on the state-dependent fatality rate, see Eq.~(\ref{eq: model equations - D}),
this condition implies that the ICU capacities must not be exceeded.
The second term in Eq.~(\ref{eq: terminal cost function}) controls
the size of the of susceptible population $S\left(T\right)$ at the
end of the epidemic. In order to approach a stable, disease-free stationary
state (``herd immunity''), the share of susceptibles on the total
population must be less than $\mathcal{R}_{0}^{-1}$ at the end of
the intervention, see Appendix~\ref{sec:Steady-state-stability}.
The term in Eq.~(\ref{eq: terminal cost function}) enforces a final
state slightly below the stability boundary (just in the stable regime),
where $0<\varepsilon\ll1$ is a small parameter. We use $\varepsilon=10^{-2}$
in the numerical simulations throughout this paper. The function
\begin{equation}
\mathcal{C}\left(x\right)=x\log{\left(x\right)}-x+1\label{eq: convex cost function}
\end{equation}
is convex on the whole domain $x\in[0,\infty)$. It appears also in
the last term of Eq.~\eqref{eq: index functional} as an intermediate
cost function, which provides an abstract measure for the total socio-economic
costs caused by the intervention. The term is minimal and zero if
no intervention is applied $\mathcal{C}\left(1\right)=\mathcal{C}'\left(1\right)=0$,
see Fig.~\ref{fig: cost functions}. The advantage of using (\ref{eq: convex cost function})
over the commonly used quadratic cost functions is that ``unphysical''
negative values of $u$ are a priori excluded. The control parameter
$\mathcal{P}$ balances between the competing objectives of minimal
disease-related deaths (first term), while attaining at the same time
a minimum number of cases to enforce $S\left(T\right)$ slightly below
the stability boundary (second term). Ramping up $\mathcal{P}$ puts
an increasing emphasis on minimizing the disease-related deaths. The
time interval $[0,T]$ of the simulation is chosen sufficiently large,
such that the results are practically independent from the chosen
final time $T$, see Tab.~\ref{tab: parameters}.

From the augmented index functional \cite{Lewis2012}
\[
\bar{\mathcal{J}}\left[u\right]=\varphi\left(\mathbf{x}\left(T\right)\right)+\int_{0}^{T}\mathrm{d}t\,\Big(\mathcal{C}\left(u\left(t\right)\right)+\boldsymbol{\lambda}\left(t\right)\cdot\left(\mathbf{F}\left(\mathbf{x}\left(t\right),u\left(t\right)\right)-\dot{\mathbf{x}}\left(t\right)\right)\Big),
\]
where $\mathbf{x}=\left(S,E,I,H,C,R,D\right)$ is the state vector,
$\dot{\mathbf{x}}=\mathbf{F}\left(\mathbf{x},u\right)$ is the dynamical
system~(\ref{eq: model equations}) and $\boldsymbol{\lambda}\left(t\right)$
is a vector of time-dependent Lagrange multipliers (also denoted as
\emph{co-state variables}) $\boldsymbol{\lambda}=\left(\lambda_{S},\lambda_{E},\lambda_{I},\lambda_{H},\lambda_{C},\lambda_{R},\lambda_{D}\right)$,
one obtains the Hamiltonian function
\begin{equation}
\mathcal{H}\left(\mathbf{x},u,\boldsymbol{\lambda}\right)=\mathcal{C}\left(u\right)+\boldsymbol{\lambda}\cdot\mathbf{F}\left(\mathbf{x},u\right).\label{eq: Hamiltonian}
\end{equation}
Following Pontryagin's maximum principle \cite{Pontryagin1962,Lewis2012},
the optimality condition reads 
\begin{equation}
\frac{\partial\mathcal{H}}{\partial u}=0\qquad\Leftrightarrow\qquad u=\exp{\left(\beta\left(\lambda_{S}-\lambda_{E}\right)\frac{IS}{N}\right)}.\label{eq: optimality}
\end{equation}
Finally, the co-state equations and the final time conditions are
obtained as 
\begin{align}
\dot{\boldsymbol{\lambda}}\left(t\right) & =-\nabla_{\mathbf{x}}\mathcal{H},\label{eq: co-state equations}\\
\boldsymbol{\lambda}\left(T\right) & =\nabla_{\mathbf{x}}\left.\varphi\left(\mathbf{x}\right)\right\vert _{T}.\label{eq: final time conditions}
\end{align}
Together with the initial conditions $\mathbf{x}\left(0\right)$,
the system $\eqref{eq: model equations}$, (\ref{eq: co-state equations})--(\ref{eq: final time conditions})
represents a nonlinear two-point boundary value problem. The full
set of equations is given in Appendix~\ref{sec: Two-point BVP and initial values}.
Numerical solutions are obtained by using Matlab's built-in routine
\texttt{bvp4c} \cite{Shampine2003} in combination with an analytic
Jacobian matrix and a step-size adaptive homotopy method, where the
control parameter $\mathcal{P}$ is gradually ramped up while always
using the result of the previous step as initialization. The procedure
is initiated from the numerical solution of the initial value problem
(\ref{eq: model equations}) without interventions, see Fig.\,\ref{fig: uncontrolled}.

\section{Results \label{sec:Results}}

\begin{figure}
\includegraphics[width=1\textwidth]{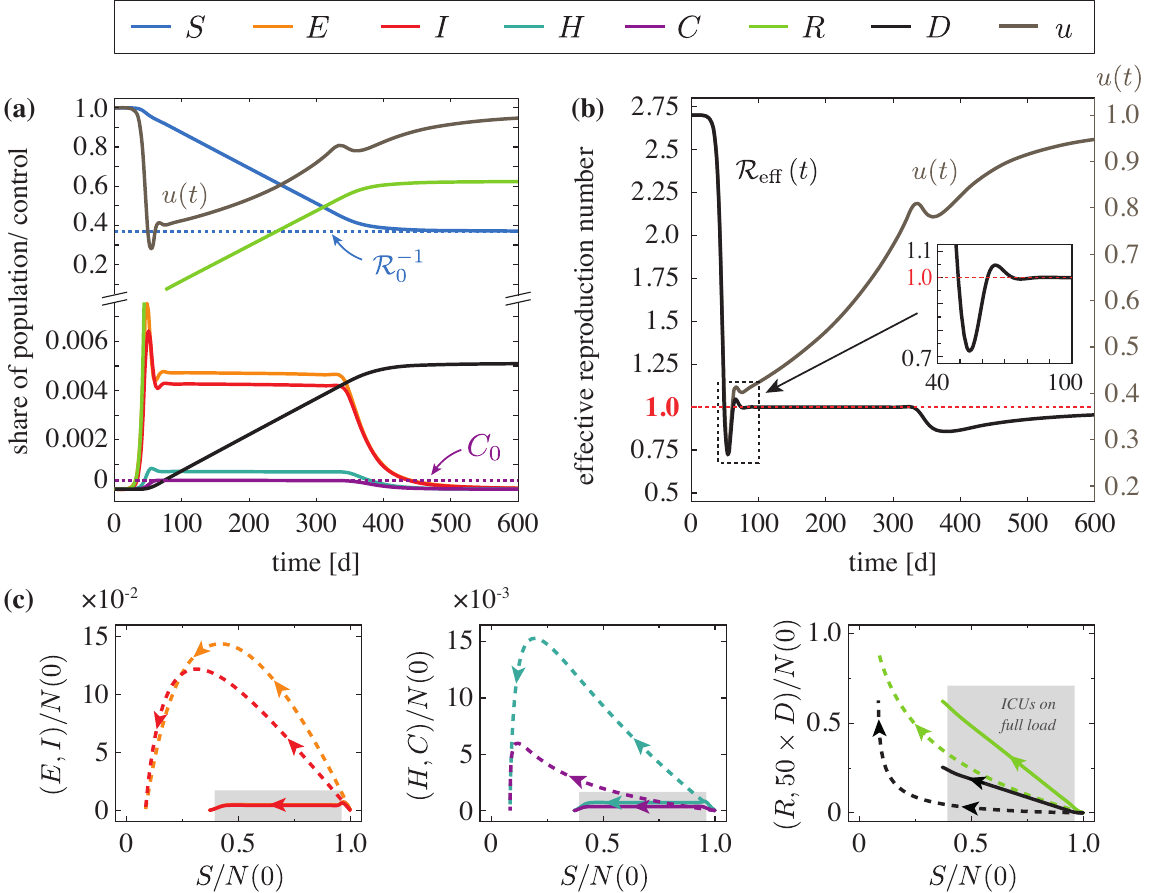}

\caption{Optimal transmission control for $C_{0}=30\thinspace000$
available ICUs. \textbf{(a)}~Temporal evolution of the optimally
controlled epidemic. The susceptible population terminates slightly
below the critical value $\mathcal{R}_{0}^{-1}$, which guarantees
herd immunity and rules out a second wave of the epidemic. Moreover,
the optimal control ensures that the available number of ICUs is not
exceeded by the critically ill: $C\left(t\right)<C_{0}$ for all $t\in[0,\infty)$.
A more detailed plot of the ICU load is given in Fig.~\ref{fig: optimal control - different number of ICUs}\,(c).
\textbf{(b)}~Effective reproduction number \eqref{eq: effective reproduction number}
corresponding to the optimally steered intervention. The optimal mean
contact reduction $u\left(t\right)$ is shown for comparison. \textbf{(c)}~Comparison
of the trajectories of the uncontrolled (dashed lines) and the optimally
controlled epidemic (solid lines) in different projections of the
state space. The arrows indicate the direction of time. The grey shaded
region highlights the critical period.}

\label{fig: optimally controlled epidemic}
\end{figure}

\subsection{Structure of the optimal intervention strategy}

\begin{figure}
\includegraphics[width=1\textwidth]{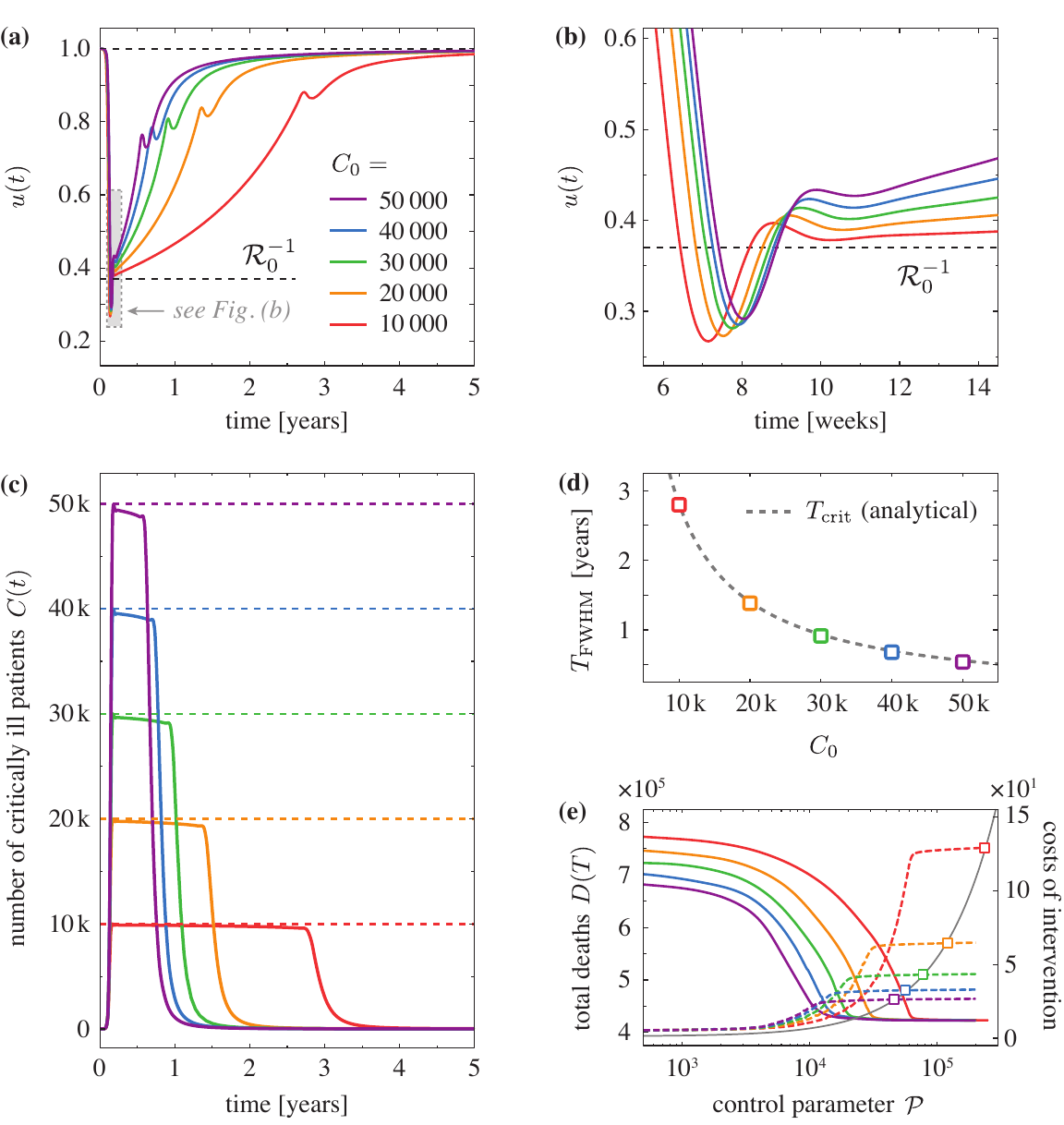}

\caption{\textbf{(a)}~Optimal time evolution of the transmission control function
$u\left(t\right)$ for different values of $C_{0}$. The value of
$C_{0}$ is color-coded. In all scenarios, the interventions start
with a strict lockdown, where $u\left(t\right)$ is reduced below
$\mathcal{R}_{0}^{-1}$ for about 10 to 12 days. This initial lockdown
is followed by a long ``critical period'' during which the measures
are gradually relaxed. The length of this period is determined by
the peak number of simultaneously critically infected $C_{0}$. Towards
the end of the intervention, a moderate tightening of the NPIs is
required. \textbf{(b)}~Same as (a), but zoomed on the region with
$u\left(t\right)<\mathcal{R}_{0}^{-1}$. \textbf{(c)}~By optimal
transmission control, the number of patients in a critical state $C$
is kept below the limiting value $C_{0}$ at all times. \textbf{(d)}~Characteristic
time span $T_{\text{FWHM}}$ of the critical period during which the
peak number of simultaneously infected must be held constant. The
dashed line shows the analytical approximation $T_{\text{crit}}$
given in Eq.~(\ref{eq: Tcrit}). \textbf{(e)}~Total number of disease-related
deaths (solid lines) and total costs of the measures (dashed lines)
at the end of the epidemic vs. the control parameter $\mathcal{P}$
(see Sec.~\ref{sec:Optimal-control}). The optimized transmission
function minimizes the number of disease-related deaths to a $C_{0}$-independent
value for $\mathcal{P}\to\infty$, but to a high cost in the case
of low $C_{0}$. The squares indicate the minimal values of $\mathcal{P}$
that guarantee $C(t)<C_{0}$ for all times.}
\label{fig: optimal control - different number of ICUs}
\end{figure}

With optimal control of the transmission rate (in the sense of Sec.~\ref{sec:Optimal-control})
via accordingly steered NPIs, the epidemic develops dramatically different
from the uncontrolled case. The whole intervention is shown in Fig.~\ref{fig: optimally controlled epidemic}
and can be structured into three phases:
\begin{enumerate}
\item The intervention begins with a strict initial ``lockdown'' that
is built up over a period of about 25 days (starting around day 25),
see Fig.~\ref{fig: optimally controlled epidemic}\,(a,\,b). The
effective reproduction number (\ref{eq: effective reproduction number})
must be held below one $\mathcal{R}_{\text{eff}}<1$ for about 13~days,
see Fig.~\ref{fig: optimally controlled epidemic}\,(b) and Fig.\,\ref{fig: optimal control - different number of ICUs}\,(b).
This strict initial intervention breaks the early exponential growth
and damps the peak number of infected such that an overshoot of the
critically ill population beyond $C_{0}$ is just barely avoided,
see Fig.\,\ref{fig: optimally controlled epidemic}\,(a) and Fig.\,\ref{fig: optimal control - different number of ICUs}\,(c).
\item The initial lockdown is followed by a long period (about 300~days
in the case of $C_{0}=30\thinspace000$), which is denoted as the
``critical period'' in the following, during which the number of
simultaneously active cases is kept approximately constant. This corresponds
to an effective reproduction number $\mathcal{R}_{\text{eff}}\approx1$,
see Fig.~\ref{fig: optimally controlled epidemic}\,(b). During
this phase, the intensive care system is constantly stressed by slightly
less than $C_{0}$ patients in a critical state. This situation
must of course be avoided in reality by all means, in particular,
since stochastic fluctuations of the case number are not included
in the deterministic model \eqref{eq: model equations} at all. During
this period, the NPIs are relaxed on a gradually increasing rate,
but initially (when the disease is not yet widespread in the population)
only very slowly, see Fig.~\ref{fig: optimally controlled epidemic}\,(b).
The duration of the critical period scales with $C_{0}^{-1}$. Further
details are discussed in Sec.~\ref{sec: critical period} below.
\item After the critical period, i.e., when the number of active cases starts
to decay, a final moderate tightening of the measures is required.
This is reflected by a notable dip in the transmission control function
and a reduction of the effective reproduction number below one, see
Fig.~\ref{fig: optimally controlled epidemic}\,(b). This final
intervention reflects the requirement to meet the herd immunity threshold
towards the end of the intervention. An unnecessarily wide overshooting
into the stable regime would result in additional infections and deaths,
see Sec.~\ref{subsec:Remarks-on-the}. Finally, the measures are
lifted on a gradually decreasing rate while the system slowly approaches
the herd immunity threshold.
\end{enumerate}
Figure~\ref{fig: optimally controlled epidemic}\,(c) shows the
trajectories of the controlled and the uncontrolled epidemic in different
state space projections. By controlling the transmission of infection,
the enormous excursion of the trajectory is prevented and the optimal
path to a stable disease-free stationary state is taken. Note that
the uncontrolled epidemic terminates far in the stable regime ($S(T)/N\left(T\right)\ll\mathcal{R}_{0}^{-1}$),
whereas in the optimally controlled case the final state is just slightly
below the stability threshold $S(T)/N\left(T\right)\apprle\mathcal{R}_{0}^{-1}$.

We point out that the optimal transmission control described above
differs from the results obtained for similar optimization problems
considered in Refs.~\cite{Alvarez2020,Ketcheson2020,Miclo2020},
which do not exhibit the distinct structural features of the intervention
(initial lockdown, critical period, final phase intervention) presented
here. A comparable result was described in Ref.~\cite{Charpentier2020},
where the intervention was divided into four different phases which
essentially coincide with our findings. Merely the lockdown was further
subdivided into a ``quick activation of a strong lockdown'' and
a ``light lockdown release.''

\subsection{Dependence on the allowed maximum number of simultaneously critical
cases \label{sec: number of ICUs}}

The state-dependent mortality rate \eqref{eq: fatality} effectively
imposes a state-constraint that strictly enforces $C<C_{0}$ for $\mathcal{P}\to\infty$,
i.e., a maximum number of simultaneously infected in a critical condition.
In principle, this allows to investigate the optimal control of other
(less extreme) scenarios, where the maximum number of simultaneously
critically infected should be held far below the number of available
ICUs (i.e., the meaning of $C_{0}$ will be reinterpreted). In this
case, the increased mortality rate $f_{1}$ is an artificial parameter
that penalizes the excess of the critically infected population over
a freely chosen threshold of $C_{0}$. By ramping up the control parameter
$\mathcal{P}$, an optimal solution with $C\left(t\right)<C_{0}$
for all $t\in[0,\infty)$ is found, that is independent of $f_{1}$.

Figure~\ref{fig: optimal control - different number of ICUs} shows
the optimal control for different values of $C_{0}$. The time course
of the optimally controlled transmission rate is qualitatively the
same for all considered values of $C_{0}$, see Fig.~\ref{fig: optimal control - different number of ICUs}\,(a,\,b).
Most notably, the time scale of the entire intervention scenario is
governed by the duration of the critical period, during which the
number of critical patients is held at $C\apprle C_{0}$, see Fig.~\ref{fig: optimal control - different number of ICUs}\,(c).
We characterize this time scale by the full width half maximum (FWHM)
time $T_{\text{FWHM}}=t_{2}-t_{1}$, where $t_{1}$ and $t_{2}>t_{1}$
are the two points in time at which the number of critically infected
equals half the allowed maximum value: $C\left(t_{1}\right)=C\left(t_{2}\right)=C_{0}/2$.
As shown in Fig.~\ref{fig: optimal control - different number of ICUs}\,(d),
the FWHM time scales inversely with the peak number of simultaneously
infected in a critical state: $T_{\text{FWHM}}\sim C_{0}^{-1}$. The
minimization of the disease-related deaths is controlled by the parameter
$\mathcal{P}$ in the terminal cost function~(\ref{eq: terminal cost function}).
Figure~\ref{fig: optimal control - different number of ICUs}\,(e)
displays the progression of the optimization routine into the targeted
optimal state (i.e., without excess of $C_{0}$) while $\mathcal{P}$
is ramped up. At a certain value of $\mathcal{P}$, which depends
on $C_{0}$, the routine reaches a plateau where both the number of
disease-related deaths as well as the total costs of the intervention
measures $\int_{0}^{T}\mathrm{d}t\,\mathcal{C}\left(u\left(t\right)\right)$
become constant. The corresponding values of $\mathcal{P}$, which
correspond to the scenario that fully avoids excess of critically
ill over $C_{0}$, are located on that plateau and are marked by square
symbols in Fig.~\ref{fig: optimal control - different number of ICUs}\,(e).
The optimized transmission function minimizes the number of disease-related
deaths to a $C_{0}$-independent value $D_{\text{min}}\left(T\right)$
for $\mathcal{P}\to\infty$, but at total cost that scales with $C_{0}^{-1}$.
An analytical estimate of the minimum attainable number of deaths
is given in Eq.~\eqref{eq: minimal number of deaths}. 

Within the present model, further reduction of disease-related deaths
below $D_{\text{min}}\left(T\right)$ can only be achieved by pharmaceutical
interventions, in particular by vaccination. The result of the $C_{0}$-independent
number of deaths at the end of the epidemic is an artifact of the
simplified modeling framework, in which a homogeneous population with
an averaged set of parameters is considered. Since the mortality rate
typically strongly depends on age and health condition, it might be
advisable to extend the model and divide the compartments into several
age or risk groups as in Refs.\,\cite{Barbarossa2020,Neher2020,Richard2020,Bonnans2020}.
The so-extended model features a matrix-valued transmission rate,
which describes the infections caused by contacts within and between
different groups, that could be further optimized by intra- and intergroup-specific
measures. This is, however, beyond the scope of this paper.

\subsection{Analysis of the critical period \label{sec: critical period}}

The numerical results shown in Fig.~\ref{fig: optimally controlled epidemic}\,(a,\,b)
indicate that during the critical period the populations $S$, $R$,
and $D$ change approximately linear, while the active cases ($E$,
$I$, $H$, $C$) are practically constant. To gain further insights,
we consider the ansatz (for $t>t^{*}$) 
\begin{align*}
S\left(t\right) & \approx N\left(0\right)-\gamma_{S}(t-t^{\ast}), & R\left(t\right) & \approx\gamma_{R}(t-t^{\ast}), & D\left(t\right) & \approx\gamma_{D}(t-t^{\ast}),
\end{align*}
where $t^{\ast}$ is a reference time that depends on the initial
conditions, $\gamma_{S}$, $\gamma_{R}$, $\gamma_{D}$ are initially
unknown rates and the infected sub-populations $\left(E,I,H,C\right)\approx\left(E^{\ast},I^{\ast},H^{\ast},C_{0}\right)$
are constant. From substituting the ansatz into the model equations~(\ref{eq: model equations}),
one obtains by a straightforward calculation analytical expressions
for the rates
\begin{align*}
\gamma_{S} & =\frac{1-c\left(1-f_{0}\right)}{\left(1-m\right)c}\gamma_{c}C_{0}, & \gamma_{R} & =\frac{1-c\left(1-mf_{0}\right)}{\left(1-m\right)c}\gamma_{c}C_{0}, & \gamma_{D} & =f_{0}\gamma_{c}C_{0},
\end{align*}
and the constants
\begin{align*}
E^{\ast} & \approx\frac{1}{\gamma_{l}}\gamma_{S}, & I^{\ast} & \approx\frac{1}{\gamma_{i}}\gamma_{S}, & H^{*} & \approx\frac{1}{\gamma_{h}}\frac{1}{cf_{0}}\gamma_{D}.
\end{align*}
The rate of new infections per day $\gamma_{S}$ during the critical
period depends only on the parameters of the disease and the maximum
capacity $C_{0}$. Note that it holds $\gamma_{S}=\gamma_{R}+\gamma_{D}$,
i.e., the number of active cases remains constant since susceptibles
become infected at the same rate on which active cases either recover
or die. The number of active cases in this \emph{dynamical equilibrium}
is a multiple of $C_{0}$: 
\begin{align*}
N_{\text{act}}^{\ast}=E^{\ast}+I^{*}+H^{\ast}+C^{\ast} & =\left(\frac{1-c\left(1-f_{0}\right)}{c\left(1-m\right)}\left(\frac{1}{\gamma_{l}}+\frac{1}{\gamma_{i}}\right)\gamma_{c}+\frac{1}{c}\frac{\gamma_{c}}{\gamma_{h}}+1\right)C_{0}.
\end{align*}
With the parameters listed in Tab.~\ref{tab: parameters}, we find
$N_{\text{act}}^{\ast}\approx28.3\,C_{0}$, i.e., one out of about
thirty infections turns critical.

\begin{figure}
\sidecaption\includegraphics{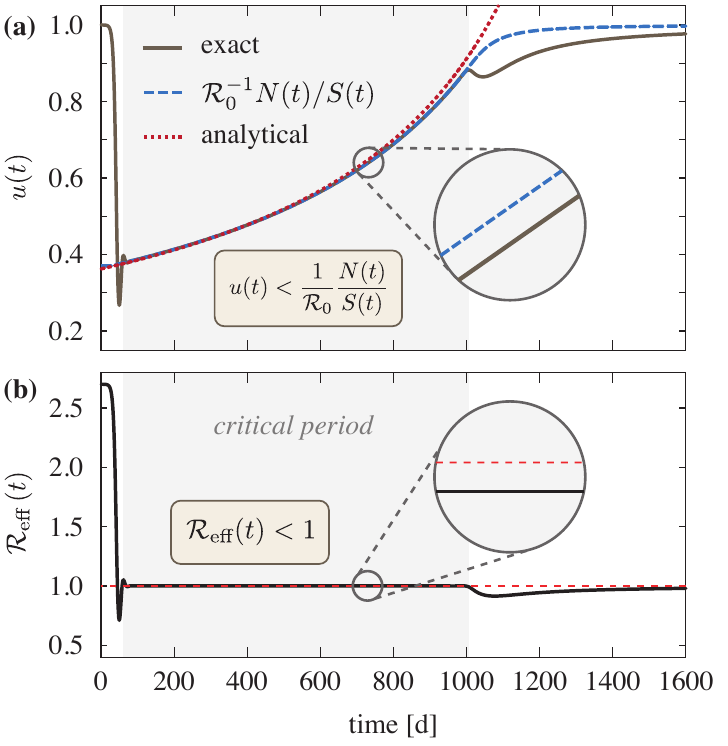}\caption{\textbf{(a)}~Analysis of the optimal mean contact reduction $u\left(t\right)$
during the critical period, where the number of simultaneously infected
must be kept constant (the plot is for $C_{0}=10\,000$). The numerically
exact result is plotted along with the stability boundary $\mathcal{R}_{0}^{-1}N\left(t\right)/S\left(t\right)$
(blue dashed line) and the analytical approximation (\ref{eq: u critical period analytical})
(red dotted line). The inset shows that the optimal control respects
the stability requirement (\ref{eq: critical period stability criterion})
during the critical period. \textbf{(b)}~Plot of effective reproduction
number $\mathcal{R}_{\text{eff}}\left(t\right)$ corresponding to
the optimal control. Throughout the critical period, $\mathcal{R}_{\text{eff}}\left(t\right)$
is kept slightly below one.\vspace{11.5ex}}

\label{fig: optimal control - critical period}
\end{figure}

Let us now come to the major results of this section. The ansatz stated
above yields an instantaneous relationship between the current value
of the transmission control function and the share of the susceptibles
on the total population $S(t)/N(t)$, which is
\begin{equation}
u\left(t\right)\approx\frac{1}{\mathcal{R}_{0}}\frac{N\left(t\right)}{S\left(t\right)}\approx\frac{1}{\mathcal{R}_{0}}\left(1-\frac{\gamma_{S}}{N\left(0\right)}(t-t^{\ast})\right)^{-1}=\left(\mathcal{R}_{0}-\left(\mathcal{R}_{0}-1\right)\frac{t-t^{\ast}}{T_{\text{crit}}}\right)^{-1}\label{eq: u critical period analytical}
\end{equation}
for a certain range of $t$ in $t^{\ast}<t<T_{\text{crit}}$ with
$T_{\text{crit}}$ defined below. Here, we approximated $N\left(t\right)\approx N\left(0\right)$
(since $\gamma_{D}\ll\gamma_{S}$). Note that Eq.~\eqref{eq: u critical period analytical}
implies $\mathcal{R}_{\text{eff}}\approx1$ during the critical period.
This approximate relation is an interesting result, as it hints that
the obtained optimal control steers the system's trajectory close
to the stability boundary. Comparison with the stability criterion
for the disease-free stationary state $\mathcal{R}_{0}<\bar{N}/\bar{S}$,
see Eq.~(\ref{eq: steady state stability criterion}), suggests that
during the critical period one must make sure that $\mathcal{R}_{\text{eff}}\left(t\right)<1$,
i.e.,
\begin{equation}
u\left(t\right)<\frac{1}{\mathcal{R}_{0}}\frac{N\left(t\right)}{S\left(t\right)}.\label{eq: critical period stability criterion}
\end{equation}
This allows to have a stable control of the number of active cases,
while the intervention measures can be gradually relaxed. Stable means
that sufficiently small fluctuations of the number of infected are
damped and do not lead to a new exponential outbreak of the epidemic.
Indeed, substituting $u\left(t\right)=\left(1+\varepsilon\right)N\left(t\right)/(\mathcal{R}_{0}S\left(t\right))$
into the model equations (\ref{eq: model equations}) yields a linear,
autonomous dynamical system (up to the state-dependent mortality rate
(\ref{eq: fatality})), which is easily seen to evolve close to a
stable dynamical equilibrium for $\varepsilon<0$ and $\left|\varepsilon\right|\ll1$,
see Appendix~\ref{sec: Stability of the Dynamical Equilibrium}.
The optimal transmission control function is shown in Fig.~\ref{fig: optimal control - critical period}
along with the analytical approximation (\ref{eq: u critical period analytical}),
the stability criterion (\ref{eq: critical period stability criterion})
and the corresponding effective reproduction number for the critical
period.

We formulate the stability criterion \eqref{eq: critical period stability criterion}
once again in a different way. Since it holds $S\left(t\right)\approx N\left(0\right)-N_{\text{cases}}\left(t\right)$,
where $N_{\text{cases}}\left(t\right)$ is the cumulative number of
cases that includes next to the active cases also the recovered and
deceased population $N_{\text{cases}}\left(t\right)=N_{\text{act}}\left(t\right)+R\left(t\right)+D\left(t\right)$,
the stability criterion (\ref{eq: critical period stability criterion})
can be written as
\begin{equation}
u\left(t\right)<\frac{1}{\mathcal{R}_{0}}\left(1-\frac{N_{\text{cases}}\left(t\right)}{N\left(0\right)}\right)^{-1}.\label{eq: criterion 1}
\end{equation}
Hence, since the optimal control depends solely on the cumulative
number of cases, it is crucial to have an accurate estimate of $N_{\text{cases}}$
at any time during the critical period. Next, we derive an upper limit
for the admissible rate of change of $u(t)$. By differentiating Eq.\,(\ref{eq: critical period stability criterion}),
using Eq.\,(\ref{eq: model equations - S}) and approximating $N\left(t\right)\approx N\left(0\right)$
as well as $I(t)\approx I^{\ast}$ (see above), we obtain
\[
\dot{u}\left(t\right)<\frac{1}{\mathcal{R}_{0}}\left(-\frac{\dot{D}\left(t\right)}{S\left(t\right)}-\frac{N\left(t\right)}{S^{2}\left(t\right)}\dot{S}\left(t\right)\right)<\frac{N\left(t\right)}{S\left(t\right)}u\left(t\right)\frac{\gamma_{i}I\left(t\right)}{N\left(t\right)}<\left(\frac{N\left(t\right)}{\mathcal{R}_{0}S\left(t\right)}\right)^{2}\frac{\mathcal{R}_{0}\gamma_{S}}{N\left(0\right)}.
\]
Using the approximation (\ref{eq: u critical period analytical}),
the rate on which the measures can be relaxed is limited by the square
of the current value of the control function. It holds
\begin{equation}
\dot{u}\left(t\right)<\frac{\mathcal{R}_{0}\gamma_{S}}{N\left(0\right)}u^{2}\left(t\right).\label{eq: criterion 2}
\end{equation}
The numerically computed optimal control obeys the criteria \eqref{eq: criterion 1}--\eqref{eq: criterion 2},
see Fig.\,\ref{fig: optimal control - critical period}, and is therefore
(weakly) stable against small perturbations. The merely weak stability
reflects the demand for minimal socio-economic costs, see Sec.\,\ref{sec:Optimal-control}.
The two rules (\ref{eq: criterion 1})--(\ref{eq: criterion 2})
for the optimal and stable steering of the transmission control function
are widely independent of the details of the current model system.
Equivalent results for a stable dynamical equilibrium with a constant
number of infected cases are easily obtained for the much simpler
SIR model.

The characteristic duration $T_{\text{crit}}$ of the critical period
is estimated from Eq.\,(\ref{eq: u critical period analytical})
and the condition $u\left(t^{\ast}+T_{\text{crit}}\right)\approx1$.
One obtains 
\begin{equation}
T_{\text{crit}}\approx\frac{N\left(0\right)}{\gamma_{S}}\left(1-\frac{1}{\mathcal{R}_{0}}\right)\propto\frac{N\left(0\right)}{C_{0}}\left(1-\frac{1}{\mathcal{R}_{0}}\right),\label{eq: Tcrit}
\end{equation}
which is in excellent agreement with the numerically obtained values
for the FWHM time plotted in Fig.~\ref{fig: optimal control - different number of ICUs}\,(d).
Finally, we estimate of the total number of disease-related deaths
from $D\left(T\right)\approx D\left(t^{*}+T_{\text{crit}}\right)\approx\gamma_{D}T_{\text{crit}}$
as
\begin{align}
D\left(T\right) & \approx N\left(0\right)\frac{\gamma_{D}}{\gamma_{S}}\left(1-\frac{1}{\mathcal{R}_{0}}\right)=N\left(0\right)\frac{\left(1-m\right)cf_{0}}{1-c\left(1-f_{0}\right)}\left(1-\frac{1}{\mathcal{R}_{0}}\right),\label{eq: minimal number of deaths}
\end{align}
which is independent of $C_{0}$, cf.~Sec.~\ref{sec: number of ICUs}
and Fig.~\ref{fig: optimal control - different number of ICUs}\,(e).

\subsection{Remarks on the final intervention phase \label{subsec:Remarks-on-the}}

\begin{figure}
\includegraphics[width=1\textwidth]{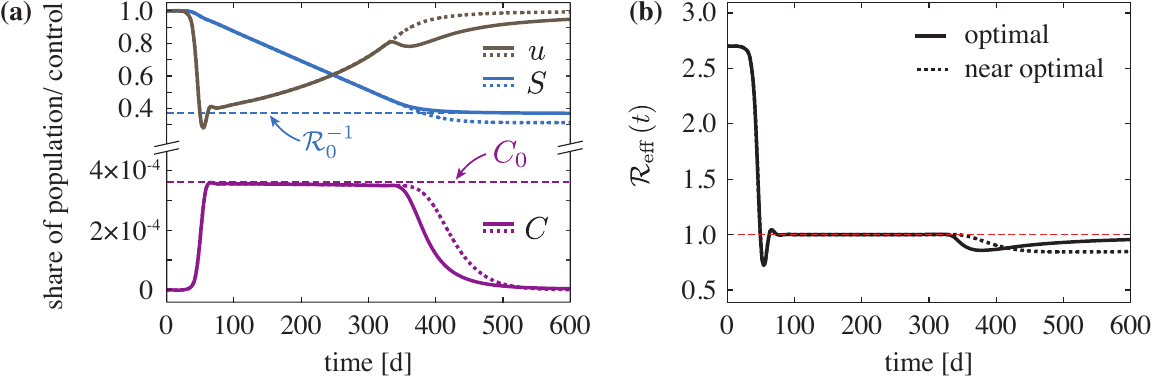}\caption{\textbf{(a)}~Comparison of the optimal (dashed) and near optimal
(dotted) control of the mean contact reduction. In the near optimal
control, the strengthening of the measures in the final phase of the
intervention is omitted. Instead, the near optimal control adheres
to the stability boundary \eqref{eq: steady state stability criterion}
and causes an overshoot of the susceptible population below the stability
threshold ($S\left(T\right)<N\left(T\right)/\mathcal{R}_{0}$), \textbf{(b)}~Plot
of the corresponding effective reproduction number.}

\label{fig: final phase}
\end{figure}

Finally, we briefly discuss the moderate tightening of the measures
in the last (third) phase of the intervention. To this end, we compare
the optimal intervention scenario with a nearly optimal control, which
lacks the last intervention phase as shown in Fig.~\ref{fig: final phase}.
In the case of nearly optimal control, the mean contact reduction
after the initial lockdown continuously follows the course of the
stability boundary \eqref{eq: steady state stability criterion},
which leads to an excess of infections beyond the required herd immunity
threshold, see Fig.~\ref{fig: final phase}\,(a). The final state
therefore is considerably further in the stable region than required.
This implies that more infections than necessary are passed through,
which results in exceeding the minimum number of deaths (not shown),
cf. Eq.~\eqref{eq: minimal number of deaths}. In order to prevent
this, the measures must be slightly tightened towards the end of the
intervention such that the number of active cases is diminished and
thus an unnecessary decrease of the susceptible population below the
herd immunity threshold is avoided.

\section{Summary and Conclusions \label{sec: Summary-and-Conclusions}}

Non-pharmaceutical measures to control the spread of infectious diseases
and to prevent a potential collapse of the health care system must
be precisely coordinated in terms of timing and intensity. Based on
well-calibrated mathematical models, the optimal intervention strategy
for specific scenarios and objectives can be computed using continuous-time
optimal control theory.

In this paper, an extended SEIR model was calibrated to reproduce
the data of the initial exponential growth phase of the COVID-19 pandemic
in Germany. Optimal control theory has been applied for the scenario
in which an effective vaccine is impossible or will never be found
and the epidemic must be controlled with purely non-pharmaceutical
measures. We have computed the optimal control of the transmission
rate that satisfies competing objectives: First, the minimization
of the disease-related deaths by strictly avoiding an overflow of
intensive care resources and, second, the suppression of a second
outbreak by establishing sufficient natural immunity at the end of
the measures. Moreover, the total costs of the intervention shall
be kept at a necessary minimum for socio-economic reasons. 

The optimal control obtained in this paper is a single-intervention
scenario that exhibits several notable features, which allow to structure
the whole intervention into three distinct phases: (i)~strict initial
lockdown, (ii)~critical period and (iii)~moderate tightening of
measures in the final phase. The obtained control differs from the
results described in related works \cite{Alvarez2020,Ketcheson2020,Miclo2020},
but is comparable to the NPI strategy presented in Ref.~\cite{Charpentier2020}.
We have shown that our optimized time-resolved NPI policy is robust
under parameter variation and developed a qualitative understanding
of its distinct phases.

The comparison of the computed optimal transmission control function
with the stability criteria (\ref{eq: criterion 1})--(\ref{eq: criterion 2})
reveals, however, that the obtained solution is in fact a tightrope
walk close to the stability boundary of the system, where socio-economic
costs and the risk of a new outbreak must be constantly balanced against
one another. Furthermore, our analysis clearly shows that the goal
of achieving herd immunity via natural infections is either extremely
expensive (in terms of socio-economic costs due to measures maintained
over a long period of time) or extremely dangerous (due to the constantly
high load on intensive care resources just below the stability limit).
Note that the values of $C_{0}$ considered in the computations are
relatively high throughout. In any case, in view of the long duration
and the enormous number of infections that this route entails, as
well as the uncertain role of sequelae and the uncertain prospects
for appropriate vaccines, it is strongly advisable to consider other
strategies, in particular the attempt to reduce the number of cases
to a level that is manageable for case tracking \cite{Meyer-Hermann2020}
or to eradicate the epidemic completely \cite{HelmholtzPositionspapier}.

\newpage

\appendix

\section{Stability Analysis of the Disease-Free Stationary State \label{sec:Steady-state-stability}}

\begin{figure}
\sidecaption\includegraphics{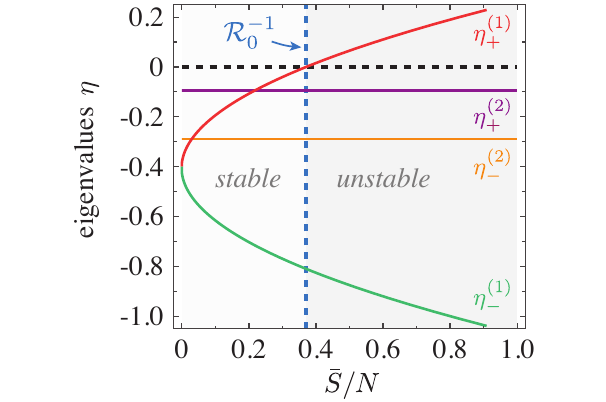} \caption{The stability of the disease-free stationary state depends on the
size of the susceptible population $\bar{S}$ and the basic reproduction
number $\mathcal{R}_{0}$. For $\bar{S}/N<\mathcal{R}_{0}^{-1}$,
the outbreak of an epidemic is suppressed by a sufficiently high degree
of herd immunity.\vspace{2.15cm}
}
\label{fig: steady state stability}
\end{figure}

Without intervention, i.e. $u=1$, the system (\ref{eq: model equations})
has a family of disease-free stationary states $\bar{\mathbf{x}}=\left(\bar{S},0,0,0,0,\bar{R},\bar{D}\right)$.
The stability of a stationary state with respect to small perturbations
$\bar{\mathbf{x}}\to\bar{\mathbf{x}}+\delta\mathbf{x}\left(t\right)$
is determined by the sign of the real parts of the eigenvalues $\eta$
of the linearized system's coefficient matrix 
\[
A\left(\bar{\mathbf{x}}\right)=\left(\begin{array}{ccccccc}
0 & 0 & -\beta\bar{S}/\bar{N} & 0 & 0 & 0 & 0\\
0 & -\gamma_{l} & \beta\bar{S}/\bar{N} & 0 & 0 & 0 & 0\\
0 & \gamma_{l} & -\gamma_{i} & 0 & 0 & 0 & 0\\
0 & 0 & \left(1-m\right)\gamma_{i} & -\gamma_{h} & \left(1-f_{0}\right)\gamma_{c} & 0 & 0\\
0 & 0 & 0 & c\gamma_{h} & -\gamma_{c} & 0 & 0\\
0 & 0 & m\gamma_{i} & \left(1-c\right)\gamma_{h} & 0 & 0 & 0\\
0 & 0 & 0 & 0 & f_{0}\gamma_{c} & 0 & 0
\end{array}\right).
\]
with $\bar{N}=\bar{S}+\bar{R}$. From the characteristic polynomial
\[
0=\chi\left(\eta\right)=\det\left(A\left(\bar{\mathbf{x}}\right)-\eta I\right),
\]
one obtains the eigenvalues 
\begin{align*}
\eta_{\pm}^{\left(1\right)} & =\frac{1}{2}\left(-\left(\gamma_{i}+\gamma_{l}\right)\pm\sqrt{\left(\gamma_{i}-\gamma_{l}\right)^{2}+4\mathcal{R}_{0}\gamma_{i}\gamma_{l}\bar{S}/\bar{N}}\right),\\
\eta_{\pm}^{\left(2\right)} & =\frac{1}{2}\left(-\left(\gamma_{c}+\gamma_{h}\right)\pm\sqrt{\left(\gamma_{c}-\gamma_{h}\right)^{2}+4c\left(1-f_{0}\right)\gamma_{c}\gamma_{h}}\right),
\end{align*}
and the threefold degenerate eigenvalue $\eta^{\left(0\right)}=0$.
Since $c\left(1-f_{0}\right)<1$, it holds $\eta_{\pm}^{\left(2\right)}<0$.
The leading eigenvalue is $\eta_{+}^{\left(1\right)}$, which is negative
for 
\begin{equation}
\bar{S}/\bar{N}<\mathcal{R}_{0}^{-1},\label{eq: steady state stability criterion}
\end{equation}
see Fig.~\ref{fig: steady state stability}. Hence, the disease-free
stationary state is unstable if the susceptible population size exceeds
a critical threshold value that is given by the inverse basic reproduction
number (\ref{eq: basic reproduction number}). For $\bar{S}/\bar{N}<\mathcal{R}_{0}^{-1}$
an epidemic outbreak is suppressed by a sufficiently high degree of
herd immunity.

\section{Dynamical Equilibrium and Stability during the Critical Period
\label{sec: Stability of the Dynamical Equilibrium}}

For the stability analysis of the dynamical equilibrium during the
critical period it is sufficient to consider the $\left(S,E,I\right)$-block
of the system \eqref{eq: model equations}, which drives the remaining
equations. At first,
\begin{align*}
\dot{S} & =-\beta u\left(t\right)\frac{IS}{N}, & \dot{E} & =\beta u\left(t\right)\frac{IS}{N}-\gamma_{l}E, & \dot{I} & =\gamma_{l}E-\gamma_{i}I,
\end{align*}
is a nonlinear and non-autonomous dynamical system. Substituting the
control function 
\[
u\left(t\right)=\left(1+\varepsilon\right)\frac{1}{\mathcal{R}_{0}}\frac{N\left(t\right)}{S\left(t\right)},
\]
yields a linear and autonomous system
\begin{align*}
\dot{S} & =-\left(1+\varepsilon\right)\gamma_{i}I, & \dot{E} & =\left(1+\varepsilon\right)\gamma_{i}I-\gamma_{l}E, & \dot{I} & =\gamma_{l}E-\gamma_{i}I.
\end{align*}
For $\varepsilon=0$, it is easily seen that $\dot{E}+\dot{I}=0$,
such that there exists a dynamical equilibrium with a constant number
of actively infected: $E^{*}+I^{*}=\text{const}$., where $E^{*}=\left(1+\gamma_{i}/\gamma_{l}\right)I^{*}$.
The corresponding susceptible population is linearly decreasing on
a rate $\gamma_{S}=\gamma_{i}I^{*}$. The stability of the dynamical
equilibrium $\left(E^{*},I^{*}\right)$ is determined by the roots
of the characteristic polynomial
\[
0=\Lambda^{2}+\left(\gamma_{l}+\gamma_{i}\right)\Lambda-\gamma_{l}\gamma_{i}\varepsilon
\]
that are easily obtained as
\[
\Lambda_{\pm}=-\frac{\gamma_{l}+\gamma_{i}}{2}\pm\sqrt{\left(\frac{\gamma_{l}+\gamma_{i}}{2}\right)^{2}+\gamma_{l}\gamma_{i}\varepsilon}.
\]
Clearly, for $\varepsilon>0$, the dynamical equilibrium becomes unstable
due to $\Lambda_{+}>0$. The stability boundary is given by $\varepsilon=0$,
on which the dynamical equilibrium exists. The optimal control obtained
in the main text drives the system slightly below the stability boundary
($\varepsilon<0$, $\left|\varepsilon\right|\ll1$), see Fig.~\ref{fig: optimal control - critical period}\,(a).
In this case it holds $\Lambda_{\pm}<0$, such that the system is
weakly stable against small perturbations, because the number of active
cases is constantly decreasing.

\section{Parameter Adjustment \label{sec:Parameter-adjustment}}

The parameters are adjusted such that the model reproduces
the data of the early exponential growth phase of the COVID-19 pandemic
in Germany. It is of course questionable to calibrate an epidemic
model to a single country, but in a scenario with extensive border
closures this seems to be justified. In the exponential growth phase
of the epidemic, all sub-populations grow exponentially with the same
rate, see Fig.~\ref{fig: uncontrolled}~(b). This observation can
be exploited to derive a series of algebraic equations (which hold
approximately in the initial phase of the epidemic) that relate all
state variables to each other. On the basis of empirical data (reported
number of cases and deaths etc.), several missing model parameters
can be directly determined from the algebraic relations. The number
of reported cases and deaths used in this study is based on the figures
provided by the Robert Koch-Institute \cite{RKIArchiv,RKIDataset}.

One starts with the ansatz 
\begin{align}
I\left(t\right) & \approx I\left(0\right)\mathrm{e}^{\Gamma t}, & S\left(t\right) & \approx N\left(0\right)\label{eq: exponential ansatz}
\end{align}
where $\Gamma$ is the initial exponential growth rate that is estimated
from reported data (see Fig.~\ref{fig: uncontrolled}\,(b)) as $\Gamma\approx0.26\,\text{d}^{-1}$
(doubling time of infections within $\Gamma^{-1}\log{\left(2\right)}\approx2.67\,\text{d}$).
Substituting Eq.~(\ref{eq: exponential ansatz}) in Eqs.~(\ref{eq: model equations - E})--(\ref{eq: model equations - I})
yields 
\[
E\left(t\right)\approx\frac{1}{\gamma_{l}}\left(\Gamma+\gamma_{i}\right)I\left(t\right)
\]
and the relation between the growth rate and $\mathcal{R}_{0}$: 
\begin{equation}
\left(1+\frac{\Gamma}{\gamma_{l}}\right)\left(1+\frac{\Gamma}{\gamma_{i}}\right)=\mathcal{R}_{0}.\label{eq: R0 gammai gammal}
\end{equation}
Note that Eq.~(\ref{eq: R0 gammai gammal}) is equivalent to the
equation for the leading eigenvalue $\eta_{+}^{\left(1\right)}$ if
the whole population is susceptible, i.e. $\Gamma=\eta_{+}^{\left(1\right)}\vert_{\bar{S}=N(0)}$
(see Appendix~\ref{sec:Steady-state-stability}). Hence, Eq.~(\ref{eq: R0 gammai gammal})
implies that the exponential growth rate $\Gamma$ changes sign at
$\mathcal{R}_{0}=1$, i.e., the epidemic recedes for $\mathcal{R}_{0}<1$.
The mean incubation period was reported to be $5.1\,\text{d}$, but
there are indications that the latency time may be shorter \cite{Lauer2020}.
Assuming the onset of infectiousness $2.5\,\text{d}$ before the onset
of symptoms, this implies an average latency period of $\gamma_{l}^{-1}=2.6\,\text{d}$,
i.e., the latency period is assumed to equal roughly half of the incubation
period. The reported values of the basic reproduction number $\mathcal{R}_{0}$
are heavily scattered. According to the Robert Koch Institute, serious
estimates range between 2.4 and 3.3 \cite{RKI_Steckbrief}. In the
following $\mathcal{R}_{0}=2.7$ shall be used, which is situated
approximately in the middle of the interval in question. From Eq.~(\ref{eq: R0 gammai gammal}),
the corresponding average infectious period is obtained as $\gamma_{i}^{-1}\approx2.35\,\text{d}.$

\begin{figure}
\sidecaption\includegraphics{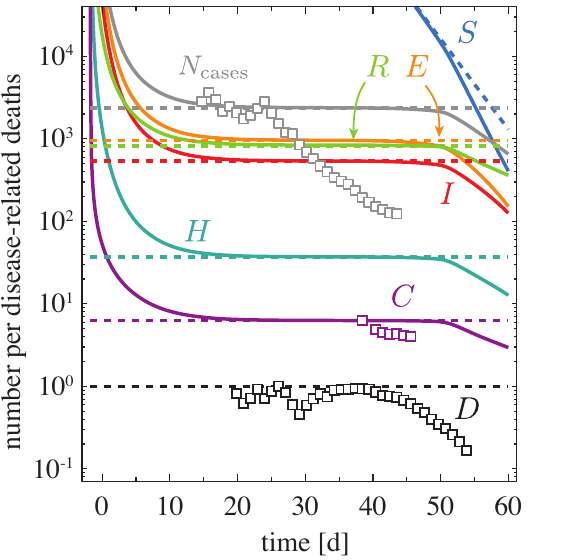}\caption{Ratio of the state variables in the initial exponential growth phase
and the number of disease related deaths. The numerically exact solution
(solid lines) is plotted along with an analytical approximation (solid
lines) that holds in the early stage of the epidemic. The corresponding
algebraic relations are used to describe ratios between different
sub-populations to facilitate the parameter adjustment. Symbols indicate
the reported number of disease-related deaths (black), estimated number
of cases (grey) and estimated ICU load (purple) of the COVID-19 pandemic
in Germany. See the text for details. During mid-March, strict social
distancing measures were implemented, that flattened the initial exponential
growth.\vspace{0.45cm}
}
\label{fig: parameter adjustment}
\end{figure}

The overall infection fatality rate of COVID-19 was estimated
as 0.66\,\% \cite{Verity2020}, such that $\left(1-m\right)cf_{0}=0.0066$.
On April 8, the Robert Koch Institute reported that a fraction of
$f_{0}=0.31$ patients in a critical state died (without ICU overflow)
\cite{RKI_8April2020}. Finally, the fraction of infected with at
most mild symptoms is estimated as $m=0.92$, such that $c=0.0213/\left(1-m\right)\approx0.266$.

Substituting the exponential ansatz (\ref{eq: exponential ansatz})
in Eqs.~(\ref{eq: model equations - H})--(\ref{eq: model equations - D}),
yields 
\begin{align*}
H\left(t\right) & \approx\frac{\left(1-m\right)}{K}\frac{\gamma_{i}}{\Gamma}\left(1+\frac{\gamma_{c}}{\Gamma}\right)I\left(t\right), & C\left(t\right) & \approx\frac{\left(1-m\right)c}{K}\frac{\gamma_{i}\gamma_{h}}{\Gamma^{2}}I\left(t\right),\\
R\left(t\right) & \approx\frac{\gamma_{i}}{\Gamma}\left(m+\frac{\left(1-m\right)\left(1-c\right)}{K}\frac{\gamma_{h}}{\Gamma}\left(1+\frac{\gamma_{c}}{\Gamma}\right)\right)I\left(t\right), & D\left(t\right) & \approx\frac{\left(1-m\right)cf_{0}}{K}\frac{\gamma_{i}\gamma_{c}\gamma_{h}}{\Gamma^{3}}I\left(t\right)
\end{align*}
with $K=1+\left(\gamma_{h}+\gamma_{c}\right)/\Gamma+\gamma_{c}\gamma_{h}\left(1-\left(1-f_{0}\right)c\right)/\Gamma^{2}$.
The analytically obtained ratio between all sub-population and deaths
(which are believed to be the most reliably reported data) are plotted
along with the corresponding numerically exact result for the initial
uncontrolled epidemic in Fig.~\ref{fig: parameter adjustment}\,(b).
The analytical results imply the relation 
\begin{equation}
D\left(t\right)/C\left(t\right)=\gamma_{c}f_{0}/\Gamma.\label{eq: exponential growth phase ratio C/D}
\end{equation}
Unfortunately, there is only little data available on the demand for
ICUs in the early phase of the epidemic. In mid-March 2020, i.e. near
the end of the initial exponential growth phase, the German Interdisciplinary
Association for Intensive Care and Emergency Medicine (DIVI) initiated
a register that reports on the availability of ICUs in Germany \cite{DIVI}.
On March 27, 687 out of 1\,160 hospitals with ICUs contributed to
the register and reported a total number of 939 COVID-19 patients
in a critical state receiving intensive care \cite{RKI_27March2020}.
At the same day, 253 disease-related deaths were reported. From the
estimated ratio $C/D\approx6.3$ (the actual number of critical patients
was estimated based on the ratio of contributing and non-contributing
hospitals as $C\approx1\thinspace586$), the average period after
which patients in a critical state either recover or die, is estimated
from Eq.~(\ref{eq: exponential growth phase ratio C/D}) as $\gamma_{c}^{-1}\approx7.5\,\text{d}$.

Finally, assuming that only $r=2/3$ of all cases have been
discovered initially and an assumed average time delay between infection
and report of cases of $\Delta t_{r}=5\,\text{d}$, the number of
actual cases is estimated from the number of reported cases as $N_{\text{cases}}^{\text{est}}\left(t\right)=r^{-1}N_{\text{cases}}^{\text{rep}}\left(t+\Delta t_{r}\right)=r^{-1}\E^{\Gamma\Delta t_{r}}N_{\text{cases}}^{\text{rep}}\left(t\right)\approx5.5\,N_{\text{cases}}^{\text{rep}}\left(t\right)$.
This yields a good agreement between the simulated number of cases
($N_{\text{cases}}=E+I+H+C+R+D$) and $N_{\text{cases}}^{\text{est}}$
before measures came into force, see Fig.~\ref{fig: uncontrolled}\,(a)
and Fig.~\ref{fig: parameter adjustment}. The average time between
infection and death $\Delta t_{d}$ can be estimated from the ratio
$N_{\text{cases}}^{\text{est}}\left(t\right)/D\left(t\right)\approx2370$
(see Fig.~\ref{fig: steady state stability}\,(b)) and $N_{\text{cases}}^{\text{est}}\left(t-\Delta t_{d}\right)=N_{\text{cases}}^{\text{est}}\left(t\right)\E^{-\Gamma\Delta t_{d}}=D\left(t\right)$
as $\Delta t_{d}\approx29.9\,\text{d}$.

\section{Two-Point Boundary Value Problem \label{sec: Two-point BVP and initial values}}

Rescaling the populations $\mathbf{x}=\left(S,E,I,H,C,R,D\right)$
subject to the dynamical system \eqref{eq: model equations} by the
initial population size $N(0)$ and using $N\left(t\right)=N\left(0\right)-D\left(t\right)$,
see Eq.~\eqref{eq: population dynamics}, we obtain the equations
of motion for the rescaled sub-populations $\mathbf{\tilde{x}}\left(t\right)=\mathbf{x}\left(t\right)/N\left(0\right)$
as
\begin{subequations}\label{eq: model equations - rescaled}
\begin{align}
\dot{\tilde{S}} & =-\beta u\left(t\right)\frac{\tilde{I}\tilde{S}}{1-\tilde{D}},\label{eq: model equations - S-1}\\
\dot{\tilde{E}} & =\beta u\left(t\right)\frac{\tilde{I}\tilde{S}}{1-\tilde{D}}-\gamma_{l}\tilde{E},\label{eq: model equations - E-1}\\
\dot{\tilde{I}} & =\gamma_{l}\tilde{E}-\gamma_{i}\tilde{I},\label{eq: model equations - I-1}\\
\dot{\tilde{H}} & =\left(1-m\right)\gamma_{i}\tilde{I}+\left(1-f(\tilde{C}/\tilde{C}_{0})\right)\gamma_{c}\tilde{C}-\gamma_{h}\tilde{H},\label{eq: model equations - H-1}\\
\dot{\tilde{C}} & =c\gamma_{h}\tilde{H}-\gamma_{c}\tilde{C},\label{eq: model equations - C-1}\\
\dot{\tilde{R}} & =m\gamma_{i}\tilde{I}+\left(1-c\right)\gamma_{h}\tilde{H},\label{eq: model equations - R-1}\\
\dot{\tilde{D}} & =f(\tilde{C}/\tilde{C}_{0})\gamma_{c}\tilde{C},\label{eq: model equations - D-1}
\end{align}
\end{subequations}where $\tilde{C}_{0}=C_{0}/N(0)$. The
co-state equations of the optimal control problem considered
in Sec.~\ref{sec:Optimal-control} for the rescaled Lagrange multipliers
$\tilde{\boldsymbol{\lambda}}\left(t\right)=N\left(0\right)\boldsymbol{\lambda}\left(t\right)$
read
\begin{align*}
\dot{\tilde{\lambda}}_{S}\left(t\right) & =L+\beta\thinspace(\tilde{\lambda}_{S}-\tilde{\lambda}_{E})\thinspace u\frac{\tilde{I}}{1-\tilde{D}},\\
\dot{\tilde{\lambda}}_{E}\left(t\right) & =L+\gamma_{l}\thinspace(\tilde{\lambda}_{E}-\tilde{\lambda}_{I}),\\
\dot{\tilde{\lambda}}_{I}\left(t\right) & =L+\beta\thinspace(\tilde{\lambda}_{S}-\tilde{\lambda}_{E})\thinspace u\frac{\tilde{S}}{1-\tilde{D}}+\gamma_{i}\left(\tilde{\lambda}_{I}-\tilde{\lambda}_{H}+m\thinspace(\tilde{\lambda}_{H}-\tilde{\lambda}_{R})\right),\\
\dot{\tilde{\lambda}}_{H}\left(t\right) & =L+\gamma_{h}\left(\tilde{\lambda}_{H}-\tilde{\lambda}_{R}+c\thinspace(\tilde{\lambda}_{R}-\tilde{\lambda}_{C})\right),\\
\dot{\tilde{\lambda}}_{C}\left(t\right) & =L+\gamma_{c}\thinspace(\tilde{\lambda}_{C}-\tilde{\lambda}_{H})+\gamma_{c}\left(f\left(\frac{\tilde{C}}{\tilde{C}_{0}}\right)+\frac{\tilde{C}}{\tilde{C}_{0}}f'\left(\frac{\tilde{C}}{\tilde{C}_{0}}\right)\right)\thinspace(\tilde{\lambda}_{H}-\tilde{\lambda}_{D}),\\
\dot{\tilde{\lambda}}_{R}\left(t\right) & =L,\\
\dot{\tilde{\lambda}}_{D}\left(t\right) & =0,
\end{align*}
with
\[
L=-\frac{u\log\left(u\right)}{1-\tilde{D}},\qquad\qquad u=\exp{\left(\beta\thinspace(\tilde{\lambda}_{S}-\tilde{\lambda}_{E})\thinspace\frac{\tilde{I}\tilde{S}}{1-\tilde{D}}\right)}.
\]
The initial conditions are taken as
\begin{align*}
\tilde{S}\left(0\right) & =1-\tilde{E}\left(0\right), & \tilde{E}\left(0\right) & =2.41\times10^{-7}, & \tilde{I}\left(0\right) & =\tilde{H}\left(0\right)=\tilde{C}\left(0\right)=\tilde{R}\left(0\right)=\tilde{D}\left(0\right)=0,
\end{align*}
and the final time conditions (\ref{eq: final time conditions}) read
\begin{align*}
\tilde{\lambda}_{S}\left(T\right) & =-\frac{1}{\varepsilon}\frac{\mathcal{R}_{0}}{1-\tilde{D}\left(T\right)}\left(1-\frac{\tilde{S}\left(T\right)}{1-\tilde{D}\left(T\right)}\right)\log\left(\frac{1}{\varepsilon}\left[1-\frac{\mathcal{R}_{0}\tilde{S}\left(T\right)}{1-\tilde{D}\left(T\right)}\right]\right),\\
\tilde{\lambda}_{E,I,H,C,R}\left(T\right) & =\frac{1}{\varepsilon}\frac{\mathcal{R}_{0}}{1-\tilde{D}\left(T\right)}\frac{\tilde{S}\left(T\right)}{1-\tilde{D}\left(T\right)}\log\left(\frac{1}{\varepsilon}\left[1-\frac{\mathcal{R}_{0}\tilde{S}\left(T\right)}{1-\tilde{D}\left(T\right)}\right]\right),\\
\tilde{\lambda}_{D}\left(T\right) & =N\left(0\right)\mathcal{P}.
\end{align*}
The choice of the initial time conditions guarantees $u\left(0\right)=1$
(no intervention) at the beginning of the scenario.


\begin{thebibliography}{10}
\providecommand{\url}[1]{{#1}}
\providecommand{\urlprefix}{URL }
\expandafter\ifx\csname urlstyle\endcsname\relax
  \providecommand{\doi}[1]{DOI~\discretionary{}{}{}#1}\else
  \providecommand{\doi}{DOI~\discretionary{}{}{}\begingroup
  \urlstyle{rm}\Url}\fi

\bibitem{Zhu2020}
Zhu, N., Zhang, D., Wang, W., Li, X., Yang, B., Song, J., Zhao, X., Huang, B.,
  Shi, W., Lu, R., Niu, P., Zhan, F., Ma, X., Wang, D., Xu, W., Wu, G., Gao,
  G.F., Tan, W.: A novel coronavirus from patients with pneumonia in {China}.
\newblock N. Engl. J. Med. \textbf{382}(8), 727--733 (2020).
\newblock \doi{10.1056/NEJMoa2001017}

\bibitem{Wu2020}
Wu, F., Zhao, S., Yu, B.e.a.: A new coronavirus associated with human
  respiratory disease in {China}.
\newblock Nature \textbf{579}, 265--269 (2020).
\newblock \doi{10.1038/s41586-020-2008-3}

\bibitem{Phua2020}
Phua, J., Weng, L., Ling, L., Egi, M., Lim, C.M., Divatia, J.V., Shrestha,
  B.R., Arabi, Y.M., Ng, J., Gomersall, C.D., Nishimura, M., Koh, Y., Du, B.:
  Intensive care management of coronavirus disease 2019 ({COVID}-19):
  {C}hallenges and recommendations.
\newblock Lancet Resp. Med.  (2020).
\newblock \doi{10.1016/s2213-2600(20)30161-2}

\bibitem{Ferguson2020}
Ferguson, N.M., Laydon, D., Nedjati-Gilani, G.e.a.: Impact of
  non-pharmaceutical interventions ({NPI}s) to reduce {COVID}-19 mortality and
  healthcare demand.
\newblock Imperial College Lond (16-03-2020)  (2020).
\newblock \doi{10.25561/77482}

\bibitem{Flaxman2020}
Flaxman, S., Mishra, S., Gandy, A., Unwin, H.J.T., Mellan, T.A., Coupland, H.,
  Whittaker, C., Zhu, H., Berah, T., Eaton, J.W., Monod, M., {Imperial College
  COVID-19 Response Team}, Ghani, A.C., Donnelly, C.A., Riley, S.M., Vollmer,
  M.A.C., Ferguson, N.M., Okell, L.C., Bhatt, S.: Estimating the effects of
  non-pharmaceutical interventions on {COVID-19} in {Europe}.
\newblock Nature  (2020).
\newblock \doi{10.1038/s41586-020-2405-7}

\bibitem{Chang2020}
Chang, S.L., Harding, N., Zachreson, C., Cliff, O.M., Prokopenko, M.: Modelling
  transmission and control of the {COVID}-19 pandemic in {Australia}.
\newblock arXiv:2003.10218  (2020)

\bibitem{Ng2020}
Ng, K.Y., Gui, M.M.: {COVID-19}: {Development} of a robust mathematical model
  and simulation package with consideration for ageing population and time
  delay for control action and resusceptibility.
\newblock Physica D: Nonlinear Phenomena \textbf{411}, 132599 (2020).
\newblock \doi{10.1016/j.physd.2020.132599}

\bibitem{Bouchnita2020}
Bouchnita, A., Jebrane, A.: A hybrid multi-scale model of {COVID-19}
  transmission dynamics to assess the potential of non-pharmaceutical
  interventions.
\newblock Chaos, Solitons \& Fractals \textbf{138}, 109941 (2020).
\newblock \doi{10.1016/j.chaos.2020.109941}

\bibitem{Jia2020}
Jia, J., Ding, J., Liu, S., Liao, G., Li, J., Duan, B., Wang, G., Zhang, R.:
  Modeling the control of {COVID}-19: {Impact} of policy interventions and
  meteorological factors.
\newblock Electronic Journal of Differential Equations (23), 1--24 (2020)

\bibitem{Kucharski2020}
Kucharski, A.J., Russell, T.W., Diamond, C., Liu, Y., Edmunds, J., Funk, S.,
  Eggo, R.M., Sun, F., Jit, M., Munday, J.D., Davies, N., Gimma, A., van
  Zandvoort, K., Gibbs, H., Hellewell, J., Jarvis, C.I., Clifford, S., Quilty,
  B.J., Bosse, N.I., Abbott, S., Klepac, P., Flasche, S.: Early dynamics of
  transmission and control of {COVID}-19: {A} mathematical modelling study.
\newblock Lancet Infect. Dis. \textbf{20}(5), 553--558 (2020).
\newblock \doi{10.1016/S1473-3099(20)30144-4}

\bibitem{Barbarossa2020}
Barbarossa, M.V., Fuhrmann, J., Heidecke, J., Vinod~Varma, H., Castelletti, N.,
  Meinke, J.H., Krieg, S., Lippert, T.: A first study on the impact of current
  and future control measures on the spread of {COVID}-19 in {Germany}.
\newblock medRxiv  (2020).
\newblock \doi{10.1101/2020.04.08.20056630}

\bibitem{Kissler2020}
Kissler, S.M., Tedijanto, C., Goldstein, E., Grad, Y.H., Lipsitch, M.:
  Projecting the transmission dynamics of {SARS-CoV-2} through the postpandemic
  period.
\newblock Science \textbf{368}(6493), 860--868 (2020).
\newblock \doi{10.1126/science.abb5793}

\bibitem{Sesterhenn2020}
Sesterhenn, J.L.: Adjoint-based data assimilation of an epidemiology model for
  the {Covid}-19 pandemic in 2020.
\newblock arXiv:2003.13071  (2020).
\newblock \doi{10.5281/zenodo.3732292}

\bibitem{Khailaie2020}
Khailaie, S., Mitra, T., Bandyopadhyay, A., Schips, M., Mascheroni, P.,
  Vanella, P., Lange, B., Binder, S., Meyer-Hermann, M.: Estimate of the
  development of the epidemic reproduction number {$R_t$} from coronavirus
  {SARS-CoV-2} case data and implications for political measures based on
  prognostics.
\newblock medRxiv  (2020).
\newblock \doi{10.1101/2020.04.04.20053637}

\bibitem{Engbert2020}
Engbert, R., Rabe, M.M., Kliegl, R., Reich, S.: Sequential data assimilation of
  the stochastic {SEIR} epidemic model for regional {COVID}-19 dynamics.
\newblock medRxiv  (2020).
\newblock \doi{10.1101/2020.04.13.20063768}

\bibitem{Dehning2020}
Dehning, J., Zierenberg, J., Spitzner, F.P., Wibral, M., Neto, J.P., Wilczek,
  M., Priesemann, V.: Inferring change points in the spread of {COVID-19}
  reveals the effectiveness of interventions.
\newblock Science  (2020).
\newblock \doi{10.1126/science.abb9789}

\bibitem{Brauner2020}
Brauner, J.M., Mindermann, S., Sharma, M., Stephenson, A.B., Gaven{\v c}iak,
  T., Johnston, D., Salvatier, J., Leech, G., Besiroglu, T., Altman, G., Ge,
  H., Mikulik, V., Hartwick, M., Teh, Y.W., Chindelevitch, L., Gal, Y.,
  Kulveit, J.: The effectiveness and perceived burden of nonpharmaceutical
  interventions against {COVID-19} transmission: {A} modelling study with 41
  countries.
\newblock medRxiv  (2020).
\newblock \doi{10.1101/2020.05.28.20116129}

\bibitem{Tsay2020}
Tsay, C., Lejarza, F., Stadtherr, M.A., Baldea, M.: Modeling, state estimation,
  and optimal control for the {US} {COVID-19} outbreak.
\newblock Sci. Rep. \textbf{10}, 10711 (2020).
\newblock \doi{10.1038/s41598-020-67459-8}

\bibitem{Tarrataca2020}
Tarrataca, L., Dias, C.M., Haddad, D.B., Arruda, E.F.: Flattening the curves:
  on-off lock-down strategies for {COVID}-19 with an application to {Brazil}.
\newblock arXiv:2004.06916  (2020)

\bibitem{Bin2020}
Bin, M., Cheung, P., Crisostomi, E., Ferraro, P., Lhachemi, H., Murray-Smith,
  R., Myant, C., Parisini, T., Shorten, R., Stein, S., Stone, L.: On fast
  multi-shot {COVID-19} interventions for post lock-down mitigation.
\newblock arXiv:2003.09930  (2020)

\bibitem{Lewis2012}
Lewis, F.L., Vrabie, D., Syrmos, V.L.: Optimal control.
\newblock John Wiley \& Sons (2012).
\newblock \doi{10.1002/9781118122631}

\bibitem{Pontryagin1962}
Pontryagin, L.S., Boltyanskii, V.G., Gamkrelidze, R.V., Mishchenko, E.F.: The
  Mathematical Theory of Optimal Processes.
\newblock John Wiley \& Sons, New York, London (1962)

\bibitem{Wickwire1977}
Wickwire, K.: Mathematical models for the control of pests and infectious
  diseases: A survey.
\newblock Theor. Popul. Biol. \textbf{11}(2), 182--238 (1977).
\newblock \doi{10.1016/0040-5809(77)90025-9}

\bibitem{Sharomi2015}
Sharomi, O., Malik, T.: Optimal control in epidemiology.
\newblock Ann. Oper. Res. \textbf{251}(1-2), 55--71 (2015).
\newblock \doi{10.1007/s10479-015-1834-4}

\bibitem{Behncke2000}
Behncke, H.: Optimal control of deterministic epidemics.
\newblock Optim. Control Appl. Meth. \textbf{21}(6), 269--285 (2000).
\newblock \doi{10.1002/oca.678}

\bibitem{Nowzari2016}
Nowzari, C., Preciado, V.M., Pappas, G.J.: Analysis and control of epidemics:
  {A} survey of spreading processes on complex networks.
\newblock IEEE Control Syst. Mag. \textbf{36}(1), 26--46 (2016).
\newblock \doi{10.1109/MCS.2015.2495000}

\bibitem{Lenhart2007}
Lenhart, S., Workman, J.T.: Optimal Control Applied to Biological Models.
\newblock Chapman \& Hall/CRC Press, Boca Raton (2007).
\newblock \doi{10.1201/9781420011418}

\bibitem{Morton1974}
Morton, R., Wickwire, K.H.: On the optimal control of a deterministic epidemic.
\newblock Adv. Appl. Probab. pp. 622--635 (1974).
\newblock \doi{10.1017/s0001867800028482}

\bibitem{Abakuks1974}
Abakuks, A.: Optimal immunisation policies for epidemics.
\newblock Adv. Appl. Probab. \textbf{6}(3), 494--511 (1974).
\newblock \doi{10.2307/1426230}

\bibitem{Zaman2008}
Zaman, G., {Han Kang}, Y., Jung, I.H.: Stability analysis and optimal
  vaccination of an {SIR} epidemic model.
\newblock Biosystems \textbf{93}(3), 240--249 (2008).
\newblock \doi{https://doi.org/10.1016/j.biosystems.2008.05.004}

\bibitem{Kar2011}
Kar, T., Batabyal, A.: Stability analysis and optimal control of an {SIR}
  epidemic model with vaccination.
\newblock Biosystems \textbf{104}(2), 127--135 (2011).
\newblock \doi{10.1016/j.biosystems.2011.02.001}

\bibitem{Zaman2009}
Zaman, G., Kang, Y.H., Jung, I.H.: Optimal treatment of an {SIR} epidemic model
  with time delay.
\newblock Biosystems \textbf{98}, 43--50 (2009).
\newblock \doi{10.1016/j.biosystems.2009.05.006}

\bibitem{Liddo2016}
Liddo, A.D.: Optimal control and treatment of infectious diseases. {The} case
  of huge treatment costs.
\newblock Mathematics \textbf{4}(2), 21 (2016).
\newblock \doi{10.3390/math4020021}

\bibitem{Gaff2009}
Gaff, H., Schaefer, E.: Optimal control applied to vaccination and treatment
  strategies for various epidemiological models.
\newblock Math. Biosci. Eng. \textbf{6}, 469--492 (2009).
\newblock \doi{10.3934/mbe.2009.6.469}

\bibitem{Hansen2011}
Hansen, E., Day, T.: Optimal control of epidemics with limited resources.
\newblock J. Math. Biol. \textbf{62}(3), 423--451 (2011).
\newblock \doi{10.1007/s00285-010-0341-0}

\bibitem{Iacoviello2013}
Iacoviello, D., Stasio, N.: Optimal control for sirc epidemic outbreak.
\newblock Comput. Methods Programs Biomed. \textbf{110}(3), 333--342 (2013).
\newblock \doi{10.1016/j.cmpb.2013.01.006}

\bibitem{Bolzoni2017}
Bolzoni, L., Bonacini, E., Soresina, C., Groppi, M.: Time-optimal control
  strategies in {SIR} epidemic models.
\newblock Math. Biosci. \textbf{292}, 86--96 (2017).
\newblock \doi{10.1016/j.mbs.2017.07.011}

\bibitem{Barro2018}
Barro, M., Guiro, A., Ouedraogo, D.: Optimal control of a {SIR} epidemic model
  with general incidence function and a time delays.
\newblock CUBO Mathematical Journal \textbf{20}(2), 53--66 (2018).
\newblock \doi{10.4067/s0719-06462018000200053}

\bibitem{Bolzoni2019}
Bolzoni, L., Bonacini, E., {Della Marca}, R., Groppi, M.: Optimal control of
  epidemic size and duration with limited resources.
\newblock Math. Biosci. \textbf{315}, 108232 (2019).
\newblock \doi{10.1016/j.mbs.2019.108232}

\bibitem{Djidjou-Demasse2020}
Djidjou-Demasse, R., Michalakis, Y., Choisy, M., Sofonea, M.T., Alizon, S.:
  Optimal {COVID}-19 epidemic control until vaccine deployment.
\newblock medRxiv  (2020).
\newblock \doi{10.1101/2020.04.02.20049189}

\bibitem{Perkins2020}
Perkins, T.A., Espa\~{n}a, G.: Optimal control of the {COVID-19} pandemic with
  non-pharmaceutical interventions.
\newblock medRxiv  (2020).
\newblock \doi{10.1101/2020.04.22.20076018}

\bibitem{Kruse2020}
Kruse, T., Strack, P.: Optimal control of an epidemic through social
  distancing.
\newblock SSRN Electronic Journal p. 3581295 (2020).
\newblock \doi{10.2139/ssrn.3581295}

\bibitem{Ketcheson2020}
Ketcheson, D.I.: Optimal control of an sir epidemic through finite-time
  non-pharmaceutical intervention.
\newblock arXiv:2004.08848  (2020)

\bibitem{Alvarez2020}
Alvarez, F.E., Argente, D., Lippi, F.: A simple planning problem for {COVID-19}
  lockdown.
\newblock Tech. rep., National Bureau of Econmic Research, Cambridge, MA
  (2020).
\newblock NBER Working Paper No. 26981

\bibitem{Bonnans2020}
Bonnans, J.F., Gianatti, J.: Optimal control techniques based on infection age
  for the study of the {COVID-19} epidemic.
\newblock HAL-02558980v2  (2020)

\bibitem{Koehler2020}
K\"{o}hler, J., Schwenkel, L., Koch, A., Berberich, J., Pauli, P.,
  Allg\"{o}wer, F.: Robust and optimal predictive control of the {COVID-19}
  outbreak.
\newblock arXiv:2005.03580  (2020)

\bibitem{Miclo2020}
Miclo, L., Spiro, D., Weibull, J.: Optimal epidemic suppression under an {ICU}
  constraint.
\newblock arXiv:2005.01327  (2020)

\bibitem{Charpentier2020}
Charpentier, A., Elie, R., Lauri\`{e}re, M., Tran, V.C.: {COVID-19} pandemic
  control: balancing detection policy and lockdown intervention under {ICU}
  sustainability.
\newblock arXiv:2005.06526  (2020)

\bibitem{Kermack1927}
Kermack, W.O., McKendrick, A.G., Walker, G.T.: A contribution to the
  mathematical theory of epidemics.
\newblock Proc. Roy. Soc. Lond. A-CONTA \textbf{115}(772), 700--721 (1927).
\newblock \doi{10.1098/rspa.1927.0118}

\bibitem{Hethcote2000}
Hethcote, H.W.: The mathematics of infectious diseases.
\newblock SIAM Review \textbf{42}(4), 599--653 (2000).
\newblock \doi{10.1137/s0036144500371907}

\bibitem{Brauer2008}
Brauer, F.: Compartmental models in epidemiology.
\newblock In: Mathematical epidemiology, pp. 19--79. Springer (2008).
\newblock \doi{10.1007/978-3-540-78911-6_2}

\bibitem{Epstein2009}
Epstein, J.M.: Modelling to contain pandemics.
\newblock Nature \textbf{460}, 687 (2009).
\newblock \doi{10.1038/460687a}

\bibitem{Rahmandad2008}
Rahmandad, H., Sterman, J.: Heterogeneity and network structure in the dynamics
  of diffusion: {Comparing} agent-based and differential equation models.
\newblock Manage. Sci. \textbf{54}(5), 998--1014 (2008).
\newblock \doi{10.1287/mnsc.1070.0787}

\bibitem{Neher2020}
Neher, R., Aksamentov, I., Noll, N., Albert, J., Dyrdak, R.: {COVID}-19
  scenarios.
\newblock Online (2020).
\newblock \urlprefix\url{https://github.com/neherlab/covid19_scenarios}.
\newblock Accessed on April 16, 2020

\bibitem{Noll2020}
Noll, N.B., Aksamentov, I., Druelle, V., Badenhorst, A., Ronzani, B.,
  Jefferies, G., Albert, J., Neher, R.: {COVID-19} scenarios: {An} interactive
  tool to explore the spread and associated morbidity and mortality of
  {SARS-CoV-2}.
\newblock medRxiv  (2020).
\newblock \doi{10.1101/2020.05.05.20091363}

\bibitem{Wilson2020}
Wilson, N., Telfar~Barnard, L., Kvalsig, A., Verrall, A., Baker, M.G., Schwehm,
  M.: Modelling the potential health impact of the {COVID}-19 pandemic on a
  hypothetical {European} country.
\newblock medRxiv  (2020).
\newblock \doi{10.1101/2020.03.20.20039776}

\bibitem{Neher2020a}
Neher, R.A., Dyrdak, R., Druelle, V., Hodcroft, E.B., Albert, J.: Potential
  impact of seasonal forcing on a {SARS-CoV-2} pandemic.
\newblock Swiss Med. Wkly. \textbf{150}, w20224 (2020).
\newblock \doi{10.4414/smw.2020.20224}

\bibitem{LloydSmith2005}
Lloyd-Smith, J.O., Schreiber, S.J., Kopp, P.E., Getz, W.M.: Superspreading and
  the effect of individual variation on disease emergence.
\newblock Nature \textbf{438}, 355--359 (2005).
\newblock \doi{10.1038/nature04153}

\bibitem{Diekmann1990}
Diekmann, O., Heesterbeek, J.A.P., Metz, J.A.J.: On the definition and the
  computation of the basic reproduction ratio {$R_0$} in models for infectious
  diseases in heterogeneous populations.
\newblock J. Math. Biol. \textbf{28}, 365--382 (1990).
\newblock \doi{10.1007/bf00178324}

\bibitem{Shampine2003}
Shampine, L.F., Gladwell, I., Shampine, L., Thompson, S.: Solving ODEs with
  Matlab.
\newblock Cambridge University Press (2003).
\newblock \doi{10.1017/cbo9780511615542}

\bibitem{Richard2020}
Richard, Q., Alizon, S., Choisy, M., Sofonea, M.T., Djidjou-Demasse, R.:
  Age-structured non-pharmaceutical interventions for optimal control of
  {COVID-19} epidemic.
\newblock medRxiv  (2020).
\newblock \doi{10.1101/2020.06.23.20138099}

\bibitem{Meyer-Hermann2020}
Meyer-Hermann, M., Pigeot, I., Priesemann, V., Sch\"{o}bel, A.: {Adaptive
  Strategien zur Eind\"{a}mmung der COVID-19-Epidemie}.
\newblock Tech. rep. (2020).
\newblock
  \urlprefix\url{https://www.mpg.de/14760567/28-04-2020_Stellungnahme_Teil_02.pdf}.
\newblock Accessed on July 13, 2020

\bibitem{HelmholtzPositionspapier}
Wiestler, O.D., Marquardt, W., Heinz, D., Meyer-Hermann, M.: {Stellungnahme der
  Helmholtz-Initiative "Systemische Epidemiologische Analyse der
  COVID-19-Epidemie" (April 13, 2020)} (2020).
\newblock
  \urlprefix\url{https://www.helmholtz.de/fileadmin/user_upload/01_forschung/Helmholtz-COVID-19-Papier_02.pdf}.
\newblock Accessed on April 17, 2020

\bibitem{RKIArchiv}
{Robert Koch-Institute}: {Archiv der Situationsberichte des Robert
  Koch-Instituts zu COVID-19}.
\newblock Online (2020).
\newblock
  \urlprefix\url{https://www.rki.de/DE/Content/InfAZ/N/Neuartiges_Coronavirus/Situationsberichte/Archiv.html}.
\newblock Accessed on April 15, 2020

\bibitem{RKIDataset}
{NPGEO Corona }: {RKI COVID19}.
\newblock Online (2020).
\newblock
  \urlprefix\url{https://npgeo-corona-npgeo-de.hub.arcgis.com/datasets/dd4580c810204019a7b8eb3e0b329dd6_0/}.
\newblock Accessed on April 16, 2020

\bibitem{Lauer2020}
Lauer, S.A., Grantz, K.H., Bi, Q., Jones, F.K., Zheng, Q., Meredith, H.R.,
  Azman, A.S., Reich, N.G., Lessler, J.: The incubation period of coronavirus
  disease 2019 ({COVID-19}) from publicly reported confirmed cases:
  {Estimation} and application.
\newblock Ann. Intern. Med. \textbf{172}(9), 577--582 (2020).
\newblock \doi{10.7326/M20-0504}

\bibitem{RKI_Steckbrief}
{Robert Koch-Institute}: {SARS-CoV-2 Steckbrief zur Coronavirus-Krankheit-2019
  (COVID-19)}.
\newblock Online (2020).
\newblock
  \urlprefix\url{https://www.rki.de/DE/Content/InfAZ/N/Neuartiges_Coronavirus/Steckbrief.html}.
\newblock Accessed on April 14, 2020

\bibitem{Verity2020}
Verity, R., Okell, L.C., Dorigatti, I., Winskill, P., Whittaker, C., Imai, N.,
  Cuomo-Dannenburg, G., Thompson, H., Walker, P.G.T., Fu, H., Dighe, A.,
  Griffin, J.T., Baguelin, M., Bhatia, S., Boonyasiri, A., Cori, A.,
  Cucunub\'{a}, Z., FitzJohn, R., Gaythorpe, K., Green, W., Hamlet, A.,
  Hinsley, W., Laydon, D., Nedjati-Gilani, G., Riley, S., van Elsland, S.,
  Volz, E., Wang, H., Wang, Y., Xi, X., Donnelly, C.A., Ghani, A.C., Ferguson,
  N.M.: Estimates of the severity of coronavirus disease 2019: {A} model-based
  analysis.
\newblock Lancet Infect. Dis. \textbf{20}(6), 669--677 (2020).
\newblock \doi{10.1016/s1473-3099(20)30243-7}

\bibitem{RKI_8April2020}
{Robert Koch-Institute}: {T\"{a}glicher Lagebericht des RKI zur
  Coronavirus-Krankheit-2019 (COVID-19): April 8, 2020}.
\newblock Online (2020).
\newblock
  \urlprefix\url{https://www.rki.de/DE/Content/InfAZ/N/Neuartiges_Coronavirus/Situationsberichte/2020-04-08-de.pdf}.
\newblock Accessed on April 14, 2020

\bibitem{DIVI}
{German Interdisciplinary Society for Intensive Care Medicine (DIVI)}: {DIVI
  Intensivregister}.
\newblock \urlprefix\url{https://www.divi.de/register/intensivregister}

\bibitem{RKI_27March2020}
{Robert Koch-Institute}: {T\"{a}glicher Lagebericht des RKI zur
  Coronavirus-Krankheit-2019 (COVID-19): March 27, 2020}.
\newblock Online (2020).
\newblock
  \urlprefix\url{https://www.rki.de/DE/Content/InfAZ/N/Neuartiges_Coronavirus/Situationsberichte/2020-03-27-de.pdf}.
\newblock Accessed on April 14, 2020

\end{thebibliography}

\end{document}